\documentclass{aa}  
\usepackage{graphicx}
\usepackage{lscape} 
\usepackage{txfonts}
\usepackage{scalerel}
\newcommand\HII{H\protect\scaleto{$II$}{1.2ex}} 
\begin{document} 

   \title{Surveys of clumps, cores, and condensations in Cygnus-X:}

   \subtitle{Searching for circumstellar disks}

   \author{Xing Pan\inst{1,2}, Keping Qiu\inst{1,2}, Kai Yang\inst{1,2}, Yue Cao\inst{1,2}, Xu Zhang\inst{1,2} 
          }

    \institute{School of Astronomy and Space Science, Nanjing University, 163 Xianlin Avenue, Nanjing 210023, P.R.China \\
              \email{kpqiu@nju.edu.cn}
         \and
              Key Laboratory of Modern Astronomy and Astrophysics (Nanjing University), Ministry of Education, Nanjing 210023, P.R.China\\
             }


 
  \abstract
   {Theories and models have suggested that circumstellar disks could channel material to the central protostar, and resist star formation feedback. Our current knowledge of the picture and role of disks around massive protostars is unclear because the observational evidence of these circumstellar disks is limited.}
   {To investigate whether disk-mediated accretion is the primary mechanism in high-mass star formation, we have established a survey of a large sample of massive dense cores within a giant molecular cloud.}
   {We used high angular resolution ($\sim 1.8''$) observations with SMA to study the dust emission and molecular line emission of about 50 massive dense cores in Cygnus-X. At a typical distance of 1.4 kpc for Cygnus-X, these massive dense cores are resolved into $\sim 2000$ au condensations. We combined the CO outflow emission and gas kinematics traced by several high-density tracers to search for disk candidates.}
   {We extracted hundreds of dust condensations from the SMA 1.3 mm dust continuum emission. The CO data show bipolar or unipolar outflow signatures toward 49 dust condensations. Among them, only 27 sources are detected in dense gas tracers, which reveals the gas kinematics, and nine sources show evidence of rotating envelopes, suggesting the existence of embedded accretion disks. The position-velocity diagrams along the velocity gradient of all rotating condensations suggest that four condensations are possible to host Keplerian-like disks.}
   {A detailed investigation of the 27 sources detected in dense gas tracers suggests that the nine disk candidates are at earlier evolutionary stages compared to the remaining 18 sources. Non-detection of rotating disks in our sample may be due to several factors, including an unknown inclination angle of the rotation axis and an early evolutionary stage of the central source, and the latter could be important, considering that young and powerful outflows could confuse the observational evidence for rotation. The detection rate of disk candidates in our sample is 1/3, which confirms that disk accretion is a viable mechanism for high-mass star formation, although it may not be the only one.}


   \maketitle
%

\section{Introduction}
Before being confirmed through observations, disks have been expected to surround forming stars of different masses since, under the conversation of angular momentum, the collapsing cloud will flatten perpendicular to the rotation axis. Numerical models \citep{2016ApJ...823...28K, 2016MNRAS.463.2553R, 2017MNRAS.471.4111H, 2018MNRAS.473.3615M} have also suggested that circumstellar disks can channel material to the central forming star, overcoming the strong feedback from the central protostar that may halt the infall of material \citep{1974A&A....37..149K, 1987ApJ...319..850W}. Circumstellar disks around low- ($M_*<2M_\odot$) and intermediate-mass ($2M_\odot<M_*<8M_\odot$) protostars have often been detected \citep[e.g.,][]{2016ApJ...828...46A, 2018ApJ...859...21A, 2017ApJ...845...44T, 2018ApJ...869L..41A}. However, the picture and role of the circumstellar disks around high-mass protostars ($M_*>8M_\odot$) are still unclear. Observational difficulties arise as they are relatively rare, distant (the typical distance is several kpc), and embedded in more crowded environments than their low-mass counterparts. In many cases, complex velocity structures and potential multiplicity can also confuse the detection \citep[e.g.,][]{2017A&A...602A..59C,2018A&A...618A..46A,2018A&A...617A.100B}. Additionally, the commonly insufficient angular resolution makes confirming the existence of disks around massive protostars a challenging task.

Searching for disks around massive stars with (sub)mm interferometers can be dated back to the late 1990s \citep{1998ApJ...505L.151Z}. Despite the aforementioned limitations, there is accumulating evidence for disks around high-mass protostars \citep{2012A&A...543A..22W, 2013A&A...552L..10S, 2014A&A...566A..73C, 2014A&A...571A..52B, 2016A&ARv..24....6B, 2017ApJ...847...58G}, thanks to instrument development. With the advent of the Atacama Large Millimeter Array (ALMA), which has sufficiently high angular resolutions ($\mathrm{<0.1"}$, corresponding to $\mathrm{\leq100\ au}$ at a typical distance of a few kpc for high-mass star-forming regions) and a high sensitivity, evidence for Keplerian rotating disks around massive protostars has been found, such as G35.20-0.74N \citep{2013A&A...552L..10S}, G35.03+0.35 \citep{2014A&A...571A..52B}, Orion source I \citep{2018ApJ...860..119G}, G17.64+0.16 \citep{2018A&A...620A..31M}, G11.92-0.61 MM1 \citep{2018ApJ...869L..24I}, and MonR2-IRS2 \citep{2020ApJ...897L..33J}. Despite the growing observational evidence of Keplerian disks (on the order of $\mathrm{10^2\sim 10^3\ au}$), there are also some studies that found no disks at similar resolutions \citep{2017ApJ...842...92G,2020ApJ...905...25G}. Considering the limited sample size of detection and non-detection cases, it is still uncertain whether the disk-mediated accretion scenario is the primary mechanism for high-mass star formation. 

With this in mind, we use the Submillimeter Array (SMA) to build a large sample of disk candidates in Cygnus-X at 1.3 mm with multiple configurations. Cygnus-X is the most massive giant molecular cloud within 3 kpc of the Sun \citep{2006A&A...458..855S}. It also has a rich collection of HII regions \citep{1991A&A...241..551W} and several OB associations \citep{2001A&A...371..675U}, making it an active star formation region in the Galaxy. Hundreds of massive dense cores (the size of a few 0.1 pc, masses of tens to hundreds of Solar masses) have been identified in the complex \citep{2007A&A...476.1243M, 2019ApJS..241....1C}. Previous observations \citep[e.g.,][]{2010A&A...524A..18B} with high angular resolutions have resolved some of these dense cores into smaller fragments on the order of 1000 au, which are termed condensations. We conducted an SMA survey of dust continuum and molecular line emission in 1.3 mm toward 48 massive dense cores identified in Cygnus-X. With a resolution of 1.8" (corresponding to $\sim2000$ au at a distance of 1.4 kpc for Cygnus-X), we find a large population of dust condensations. As a part of our project, Surveys of Clumps, CorEs, and CoNdenSations in CygnUS-X (CENSUS, PI: Keping Qiu), this paper focuses on the outflow structures and gas kinematics around the dust condensations. Our SMA survey reveals many CO outflows originating in condensations. High-density tracers (e.g., $\mathrm{CH_3CN}$, $\mathrm{CH_3OH}$ and $\mathrm{H_2CO}$ lines) have allowed us to further investigate the molecular gas kinematics of the condensations.
This paper is structured as follows. Section \ref{sec:observations} summarizes the observations and data reduction. Section \ref{sec:results} presents the images of 1.3 mm dust continuum, outflows, and the velocity field traced by selected molecular lines along with the derived gas temperature. In Section \ref{sec:discussion}, we provide the detection rate and its implications for high-mass star formation. Summaries are presented in Section \ref{sec:summary}.

\section{Observations and data reduction}\label{sec:observations}
We first chose massive dense cores (MDCs) identified by \citet{2007A&A...476.1243M} as the targets of the SMA high-resolution observations. Because the region toward Cygnus OB2 was not covered by \citet{2007A&A...476.1243M}, we then added eight dust continuum peaks seen in the JCMT 850 $\mu$m maps of the OB2 region \citet{2019ApJS..241....1C}. Since we are searching for disk candidates toward dust condensations in this work, we excluded the SMA data that detected no dust condensations or for which the 1.3 mm continuum emission was heavily contaminated by the free-free emission. In the end, the SMA observations presented in this work were done with 31 single-pointing fields and two mosaic fields, covering 48 MDCs identified by \citet{2021ApJ...918L...4C} (see Table \ref{tab:mdcpara}). It is also worth noting that all these MDCs were included in a VLA study of compact radio sources by \citet{2022ApJ...927..185W}, which is also part of the CENSUS project. Detailed information about our SMA observations is presented in Table \ref{tab:obspara}.

We calibrated the raw data with the IDL superset MIR\footnote{\url{https://lweb.cfa.harvard.edu/~cqi/mircook.html}}, and exported the calibrated visibilities into CASA \citep{2007ASPC..376..127M} for joint imaging. All images were made using a Briggs weighting of 0.5 \citep{1995AAS...18711202B} to balance resolution and sensitivity. The resulting synthesized beams are $\sim1.8''$ in the final maps. During our observations, the SMA correlator was being upgraded from the ASIC to SWARM. Also considering the archive observations, the data used in this project had total bandwidths ranging from 4 GHz to 16 GHz, and the spectral resolution also varied. However, all the high-density tracers relevant to this work and the outflow tracers, CO (2-1), $^{13}\mathrm{CO}$ (2-1), and SiO (5-4), were all covered by each observation. We also smoothed all the data into a uniform-frequency resolution of 812.5 kHz ($\sim1$ km s$^{-1}$ at 230 GHz) during imaging.
\begin{table*}[!ht]
\caption{List of observational parameters}
\label{tab:obspara}
\begin{tabular}{clcllll}
\hline
\hline
Observation & Array & $\mathrm{N_{Ant}}$\tablefootmark{a} & Bandpass & Gain & Flux & Observation \tablefootmark{b} \\
Date & Configuration &  & Calibrator & Calibrator & Calibrator & Field\\
\hline
2011/05/31 & Compact & 7 & 3C279 & MWC349a & Titan & 33 \\
2011/06/03 & Compact & 8 & 3C279 & MWC349a & Titan/Uranus & 33 \\
2011/06/27 & Subcompact & 7 & 3C279 & MWC349a & Titan/Uranus & 33 \\
2012/07/08 & Compact & 6 & 3C279 & MWC349a & Uranus & 32 \\
2012/08/03 & Subcompact & 6 & 3C279 & MWC349a & Titan/Mars & 32 \\
2012/08/16 & Subcompact & 5 & 3C279 & MWC349a & Saturn & 32 \\
2015/06/14 & Compact & 7 & 3C84 & MWC349a & Neptune & 1/2/3/4 \\
2015/06/16 & Compact & 7 & 3C84 & MWC349a & Neptune/Uranus &  21/22/23/24 \\
2015/06/18 & Compact & 7 & 3C454.3 & MWC349a & Neptune &  6/7/8 \\
2015/07/05 & Compact & 6 & 3C279 & MWC349a & Neptune &  1/2/3/4 \\
2015/07/06 & Compact & 6 & 3C454.3 & MWC349a & Neptune &  6/7/8 \\
2015/07/07 & Compact & 6 & 3C454.3 & MWC349a & Neptune &  9/10/11/12/17 \\
2015/07/14 & Compact & 6 & 3C279 & MWC349a & Uranus &  13/14/25 \\
2015/07/18 & Compact & 6 & 3C279 & MWC349a & Neptune &  9/10/11/12/17 \\
2015/07/19 & Compact & 6 & 3C279 & MWC349a & Neptune &  13/14/16/25 \\
2015/07/26 & Compact & 6 & 3C279/3C454.3 & MWC349a & Titan &  5/18/19/20 \\
2015/08/02 & Compact & 5 & 3C454.3 & MWC349a & Neptune &  21/22/23/24 \\
2015/08/09 & Compact & 6 & 3C454.3 & MWC349a & Neptune &  21/22/23/24 \\
2015/09/18 & Extended & 7 & 3C454.3 & MWC349a & Neptune &  13/14/16/23/24/25 \\
2015/09/24 & Extended & 7 & 3C454.3 & MWC349a & Neptune/Uranus &  17/18/19/20/21/22 \\
2015/09/25 & Extended & 7 & 3C454.3 & MWC349a & Neptune/Uranus &  8/9/10/11/12 \\
2015/10/02 & Extended & 7 & 3C454.3 & MWC349a & Neptune &  1/2/3/4/5/6/7 \\
2015/10/03 & Extended & 7 & 3C454.3 & MWC349a & Neptune/Uranus &  8/9/10/11/12 \\
2015/10/07 & Extended & 7 & 3C454.3 & MWC349a & Neptune/Uranus &  8/9/10/11/12 \\
2015/10/21 & Subcompact & 6 & 3C454.3 & MWC349a & Uranus &  1/2/3/9/10/15 \\
2015/10/22 & Subcompact & 6 & 3C454.3 & MWC349a & Uranus &  8/13/14/20/23 \\
2015/10/28 & Subcompact & 7 & 3C454.3 & MWC349a & Uranus &  11/12/15/16 \\
2015/11/01 & Subcompact & 7 & 3C454.3 & MWC349a & Neptune &  1/2/3/4/5/6/7/8/9/10 \\
2015/11/04 & Subcompact & 7 & 3C454.3 & MWC349a & Uranus &  13/14/17/18/19/\\
& & & & & & 20/21/22/23/24/25 \\
2016/07/15 & Very Extended & 8 & 3C454.3 & MWC349a & Neptune &  9/16/22/32 \\
2016/07/29 & Very Extended & 8 & 3C273 & MWC349a & Titan &  33 \\
2016/08/16 & Subcompact & 6 & 3C279 & MWC349a & Neptune &  15/19 \\
2016/08/25 & Subcompact & 8 & 3C279/3C454.3 & MWC349a & Neptune &  26/27/28/29/30/31 \\
2016/10/06 & Extended & 7 & 3C454.3 & MWC349a & Neptune &  33 \\
2016/11/22 & Extended & 8 & 3C84 & MWC349a & Neptune &  1/2/3/4/5/8/9/10/11/12 \\
2016/11/23 & Extended & 8 & 3C84 & MWC349a & Neptune &  17/18/20/21/23/25 \\
\hline
\end{tabular}
\tablefoot{\\
\tablefoottext{a}{Number of available antennae in each observation.}\\
\tablefoottext{b}{ID of Field from Table \ref{tab:mdcpara} covered in each observation.}
}
\end{table*}

\section{Results} \label{sec:results}
\subsection{Dust condensations and outflows}\label{subsec:outflow}
\begin{figure*}[!htp]
\centering
\includegraphics[width=15.5cm]{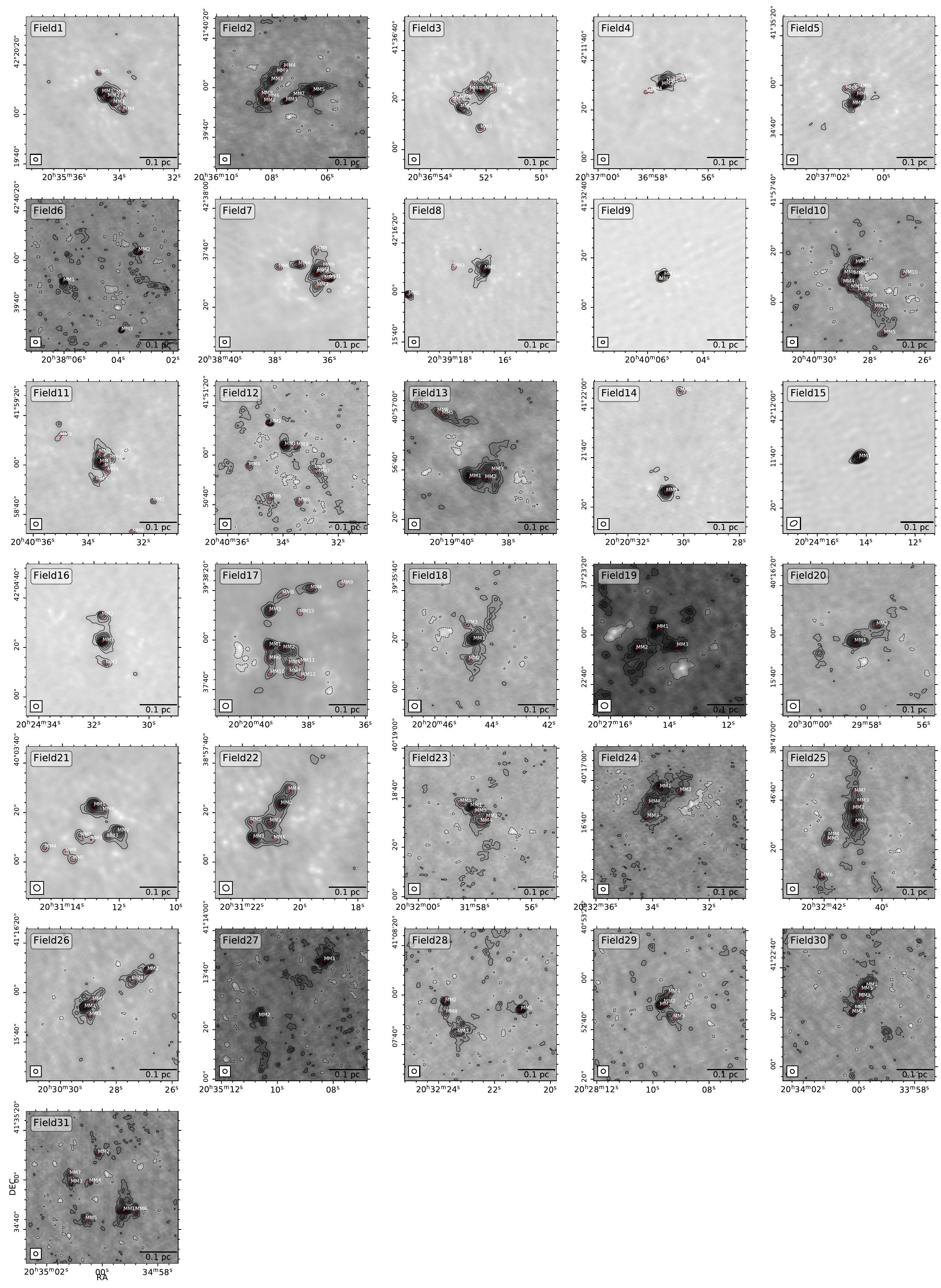}
\caption{SMA 1.3 mm continuum emission of sources in our sample. The contour levels are (-3, 3, 6, 9, 12, 18, 24, 36, 48, 60)$\times\sigma$, where $\sigma$ is the rms of the dust continuum. The synthesized beam of each image is shown in the bottom left. The red plus represents the dust condensation extracted by \citet{2021ApJ...918L...4C}.}
\label{fig:1.3mmContMap001}
\end{figure*}

The 1.3 mm continuum maps of sources in our sample are shown in Fig \ref{fig:1.3mmContMap001} (Fig \ref{fig:1.3mmContMap003} and Fig \ref{fig:1.3mmContMap004} show continuum maps of the two mosaic fields). With a typical resolution of $\sim$1.8" ($\sim$2000 au), these massive dense cores (MDCs, $\sim0.1$ pc) are resolved into dust condensations ($\sim0.01$ pc). Most fields harbor a few bright and compact condensations, except Field15, which appears to be dominated by a single source. Some fields, such as Field2 and Field17, exhibit high levels of fragmentation. \citet{2021ApJ...918L...4C} employed ``\emph{getsources}''\footnote{A powerful multi-scale, multi-wavelength source extraction algorithm \citet{2012A&A...542A..81M}. ``\emph{getsources}'' is designed for extracting dense structures in star-forming regions, decomposing images into components and measuring their properties by removing the large-scale background.} to identify condensation from these MDCs. Applying the criterion that the emission peak must exceed 5$\sigma$ of the image noise levels, the robust detections of dust condensations are about 200. The full width at half maximum (FWHM) diameters of these condensations range from 0.008 to 0.05 pc.

\begin{figure*}[!h]
    \centering
    \includegraphics[width=15.5cm]{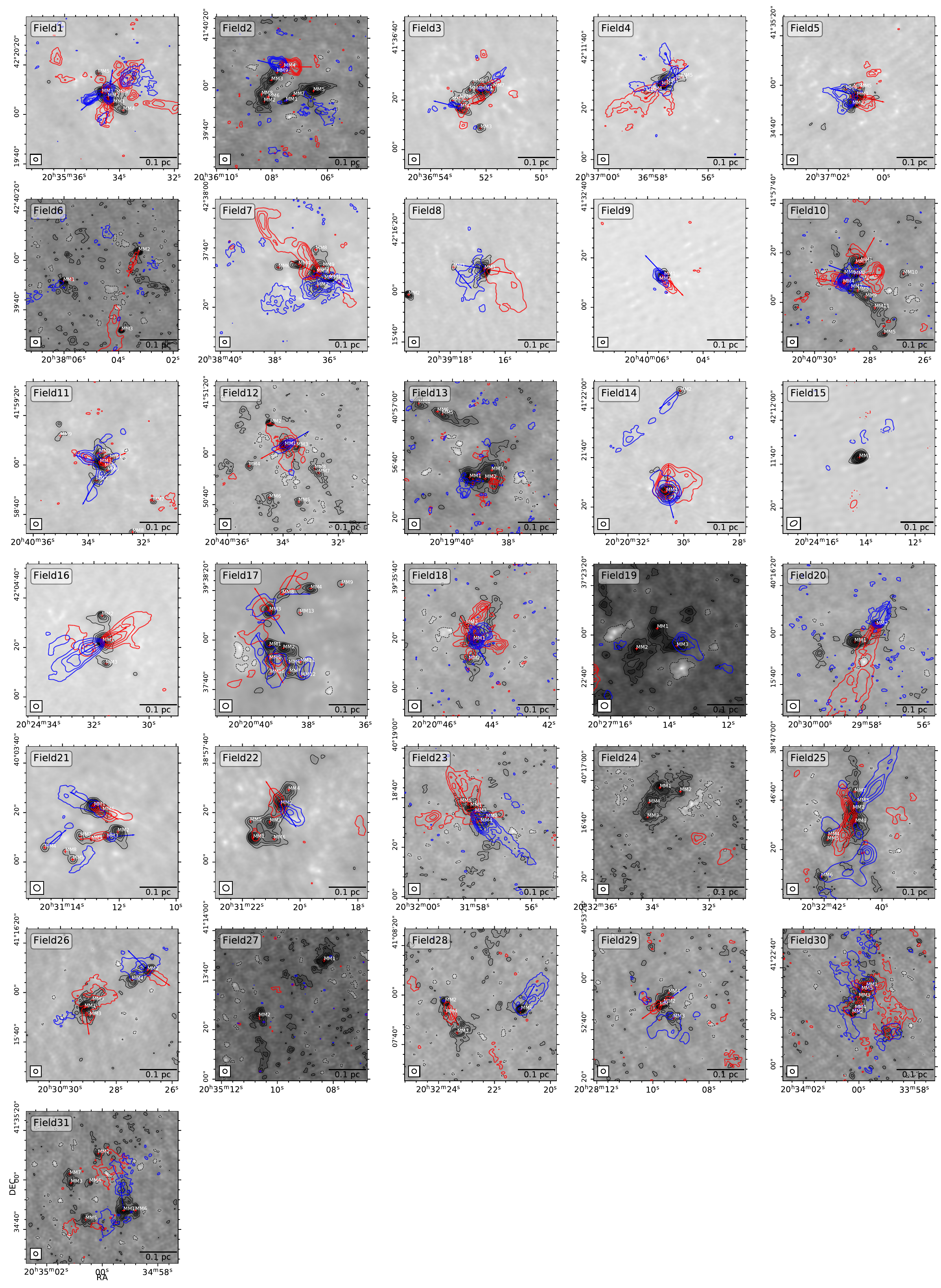}
    \caption{Molecular outflows detected in $\mathrm{CO\ (2-1)}$ in some sources. The blueshifted and redshifted outflows of $\mathrm{CO\ (2-1)}$ are plotted in blue and red contours, respectively. The contour levels are (-3, 3, 6, 9, 12, 15, 18, 24, 36, 48, 60)$\times\sigma$, where $\sigma$ is the rms of the the integrated CO emission. The blue and red arrows mark the outflow axes identified in the region. The grayscale images and black contours show SMA 1.3mm continuum emission. The contour levels are the same as in Fig \ref{fig:1.3mmContMap001}. The extracted dust condensation is marked by a black cross. }
    \label{fig:OutflowOverCont001}
\end{figure*}

To search for disk candidates, first we have to find condensations located at the geometrical center of a bipolar outflow or connected to a unipolar outflow, since accretion disks are expected to be associated with jets or outflows ejected along their rotation axes \citep{2007prpl.conf..197C}. CO emission lines usually trace the molecular outflowing gas. The velocity-integrated intensity maps of $\mathrm{^{12}CO\ 2-1}$ toward sources in our sample are shown in Fig \ref{fig:OutflowOverCont001} (Fig \ref{fig:OutflowOverContdr15} and Fig \ref{fig:OutflowOverContDR21OH} show outflow structures of the two mosaic fields). We use red and blue arrows to depict redshifted and blueshifted outflow axes, respectively. $\mathrm{SiO\ 5-4}$ emission also help us confirm the outflow axis. The details of the $\mathrm{SiO\ 5-4}$ emission are presented in \citet{yang2023subm}. In total, we find 49 dust condensations associated with outflows (hereafter “outflow-associated” condensations) and expect to find disk candidates among them.

\subsection{High-density tracers}\label{subsec:densegas}
The most reliable criteria of disk identification are based on the velocity field, which requires us to study the kinematics of these 49 outflow-associated condensations. We looked through all the detected molecular line emission from outflow-associated condensations and found that only 27 of these condensations have high-density tracers to investigate the kinematics. These 27 condensations are labeled as “dense-gas-traced.” For the other 22 outflow-associated condensations, we cannot study their kinematics, and therefore we leave them out in the following analyses.
\begin{table}[!ht]
\centering
\caption{Information on the spectral lines }
\label{tab:lineinfo}
{
\begin{tabular}{cccc}
\hline
\hline 
Line & Trasnition & Frequency & $\mathrm{E_{up}}$\\
ID & & GHz & K \\
\hline
1 & $\mathrm{C^{18}O\ 2-1}$ & 219.560358 & 15.8 \\
2 & $\mathrm{^{13}CO\ 2-1}$ & 220.398684 & 15.9 \\
3 & $\mathrm{^{12}CO\ 2-1}$ & 230.538000 & 16.6 \\
4 & $\mathrm{SiO\ 5-4}$ & 217.104919 & 31.3 \\
5 & $\mathrm{DCN\ 3-2}$ & 217.238538 & 20.9 \\
6 & $\mathrm{SO\ 6_5-5_4}$ & 219.949442 & 35.0 \\
7 & $\mathrm{H_2CO\ 3_{0,3}-2_{0,2}}$ & 218.222195 & 21.0 \\
8 & $\mathrm{H_2CO\ 3_{2,2}-2_{2,1}}$ & 218.475632 & 68.1 \\
9 & $\mathrm{H_2CO\ 3_{2,1}-2_{2,0}}$ & 218.760066 & 68.1 \\
10 & $\mathrm{CH_3OH\ 4_{2,2}-3_{1,2}}E$ & 218.440050 & 45.4 \\
11 & $\mathrm{CH_3OH\ 8_{0,8}-7_{1,6}}E$ & 220.078561 & 96.6 \\
12 & $\mathrm{CH_3OH\ 8_{-1,8}-7_{0,7}}E$ & 229.758756 & 89.1 \\
13 & $\mathrm{CH_3CN\ 12_0-11_0}$ & 220.747261 & 68.9 \\
14 & $\mathrm{CH_3CN\ 12_1-11_1}$ & 220.743011 & 76.0 \\
15 & $\mathrm{CH_3CN\ 12_2-11_2}$ & 220.730260 & 97.4 \\
16 & $\mathrm{CH_3CN\ 12_3-11_3}$ & 220.709016 & 133.2 \\
17 & $\mathrm{CH_3CN\ 12_4-11_4}$ & 220.679287 & 183.2 \\
18 & $\mathrm{CH_3CN\ 12_5-11_5}$ & 220.641084 & 247.4 \\
19 & $\mathrm{CH_3CN\ 12_6-11_6}$ & 220.594423 & 325.9 \\
20 & $\mathrm{CH_3CN\ 12_7-11_7}$ & 220.583550 & 418.6 \\
21 & $\mathrm{CH_3CN\ 12_8-11_8}$ & 220.475807 & 525.6 \\
\hline\\

\end{tabular}
}
\end{table}

\begin{figure*}[!ht]
    \centering
    \includegraphics[width=16cm]{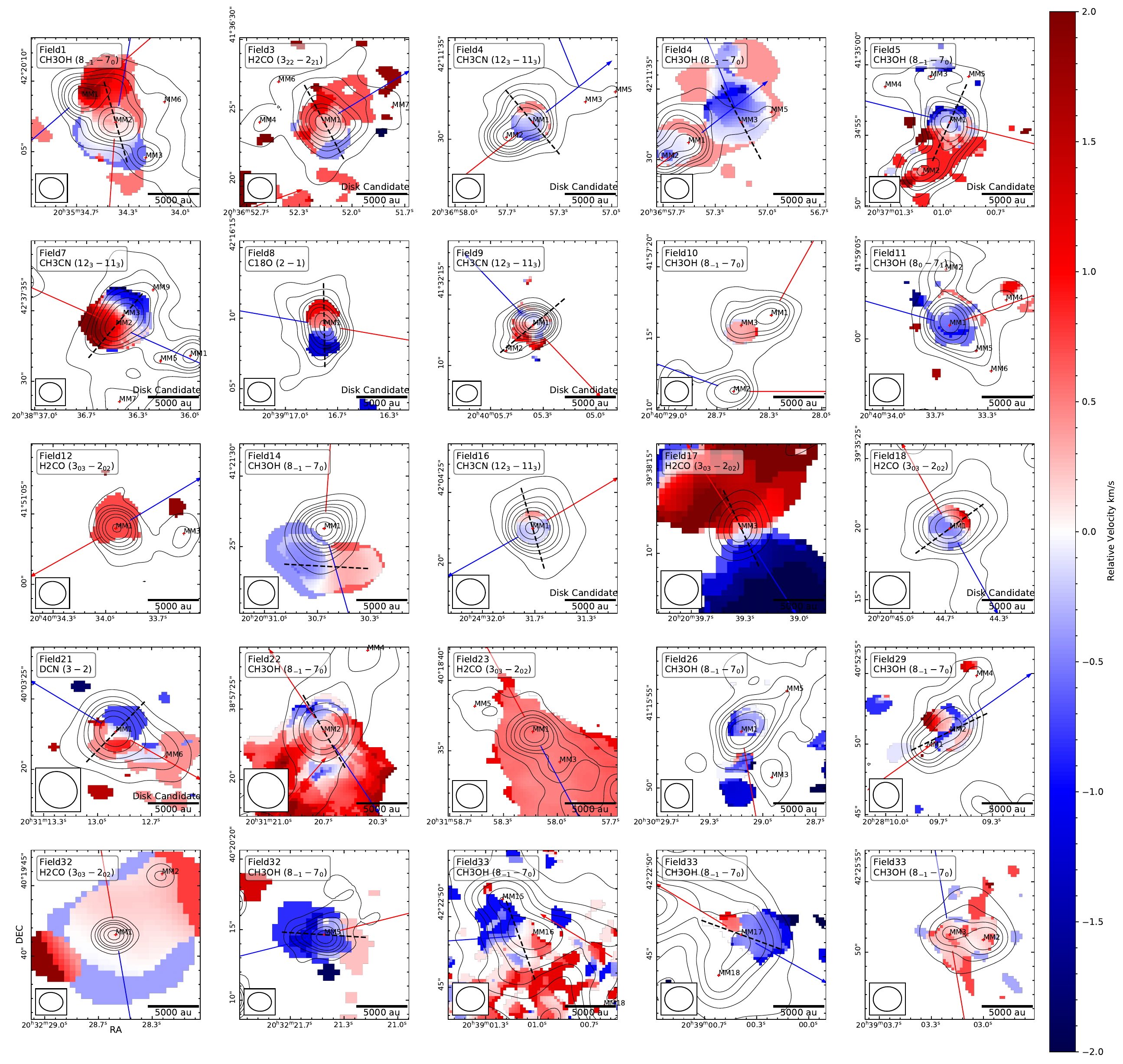}
\caption{Intensity-weighted mean velocity maps of representative molecular line emission toward all dense-gas-traced sources, overlaid with the SMA 1.3mm continuum emission contours, the same as in Fig \ref{fig:1.3mmContMap001}. The background color map shows the relative velocity to the systemic velocity of the source. The blue and red arrows represent the blueshifted and redshifted CO outflow axis, respectively. The dashed black lines indicate the velocity gradient orientation inferred from molecular line kinematics. Each panel is labeled with the species name and the excitation energy (in Kelvin) of the upper level of the transition. The label “disk candidate” on the bottom right marks the disk candidates in our sample.}
    \label{fig:velo-dense001}
\end{figure*}

We detect a series of molecular lines tracing high-density gas for each dense-gas-traced condensation, including transitions from $\mathrm{CH_3CN}$, $\mathrm{CH_3OH}$, $\mathrm{H_2CO}$, and DCN. The information on these transitions is listed in Table \ref{tab:lineinfo}. $\mathrm{CH_3CN}$ lines have proven to be a great dense gas and disk tracer. However, we only detect $\mathrm{CH_3CN}$ emission in a few sources. Considering the various evolutionary stages and physical and chemical properties of condensations in our sample, it is difficult to find a uniform disk tracer for all the condensations. Consequently, we employed different species to investigate the kinematics in our study. The intensity-weighted mean velocity (first-moment) maps of the representative molecular line toward dense-gas-traced condensations are shown in Fig \ref{fig:velo-dense001}. We chose the molecular spectral lines to show the dense gas kinematics according to the following criteria: (i) the molecular line should show compact emission around the central source of the outflow, which ensures the emission mainly come from the circumstellar disk or surrounding envelope; (ii) if several dense gas tracers all exhibited a velocity gradient in a similar direction, we chose the molecular line whose emission exhibited the clearest velocity gradient. In most cases, we had seven molecular line emission from five species to investigate the kinematics of these dense-gas-traced condensations. The high-excitation transitions of $\mathrm{CH_3CN}$ and $\mathrm{CH_3OH}$ are common dense gas tracers to search for rotating disks or “toroids.” The line emission from low energy levels of some molecular species, $\mathrm{DCN}$ and $\mathrm{H_2CO}$, can also trace the velocity fields around condensation. In some cases \citep[e.g.,][Field8 MM1 in our sample]{2020ApJ...905..162T, 2022ApJ...933..178C}, $\mathrm{C^{18}O}$ (2-1) emission can also exhibit a velocity gradient as the result of a rotating disk or envelope. 

\subsection{Identification of disk candidates}\label{subsec:diskID}
The rotating disk candidates can be identified by two simple criteria following \cite{2017A&A...602A..59C}: (i) the position angle of the observed velocity gradient should be within 20$^\circ$ orthogonal to the CO outflow axis and (ii) the relative velocity with respect to the systemic velocity of the source should increase toward the center in the position-velocity (PV) plots. The position angles of the outflow axis and velocity gradient are given in Table \ref{tab:outflowassociated}. We employ relatively loose constraints (20 degrees) on the relative angles between the outflow axis and velocity gradient for several reasons. First, the position angles of the outflow axis and velocity gradient are generally defined by eyes. The complex velocity fields and asymmetries in the outflow emission can bring large uncertainties to the measurement of the position angles. Second, in some cases the dust condensations are clustered, making it difficult to confirm which one is the central source of the outflow. Among the 27 dense-gas-traced condensations, only nine display velocity gradients roughly perpendicular to the outflow axis. We label these condensations as disk candidates.

\begin{figure}[!h]
    \centering
    \includegraphics[width=8.5cm]{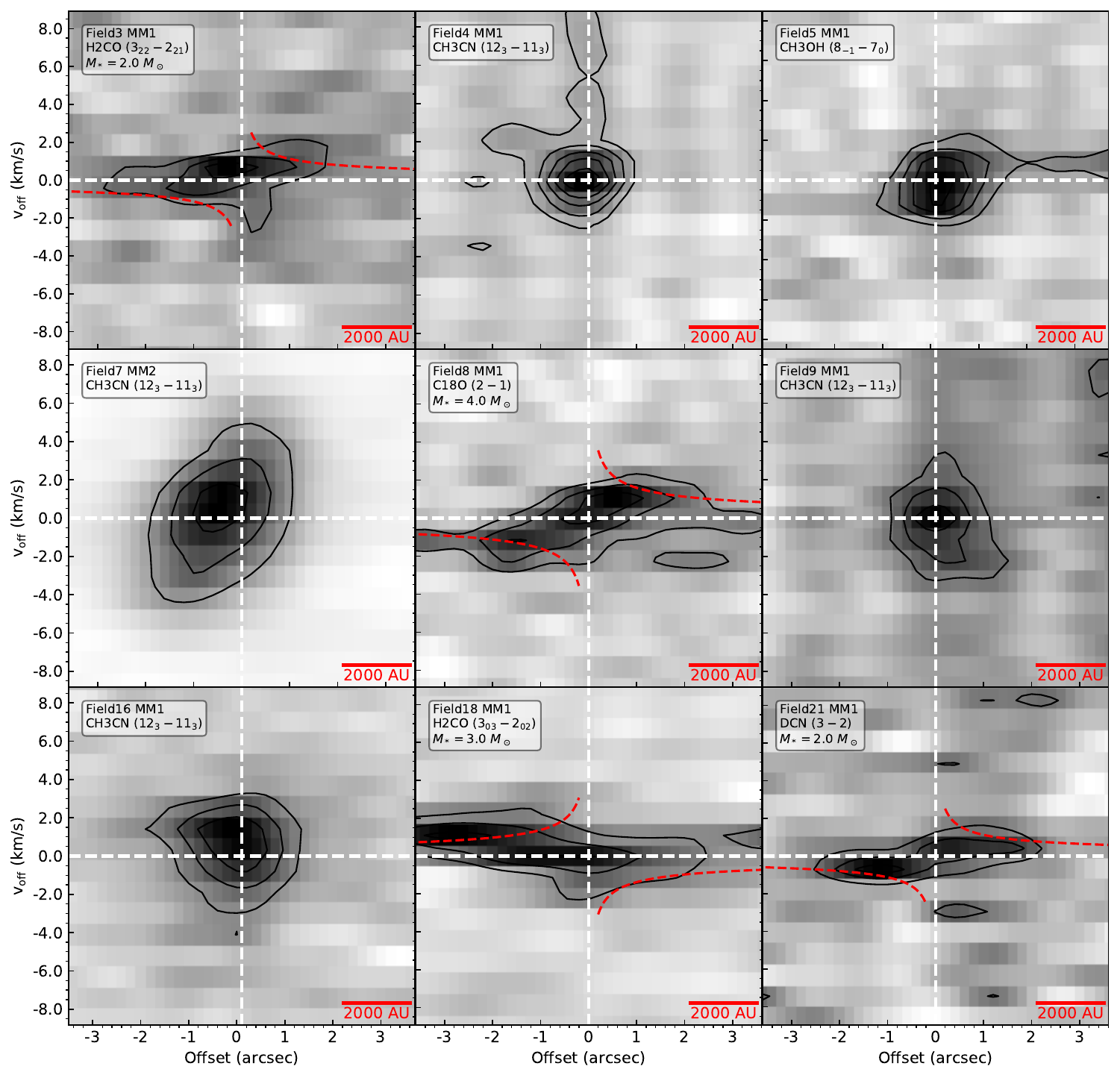}
    \caption{\label{fig:pvmap}PV diagram of spectral lines toward the disk candidates, cut along the dashed black lines shown in Fig \ref{fig:velo-dense001}. The dashed red lines show the Keplerian rotation curves for central masses labeled in the upper left. The black contours correspond to 3, 5, 7, 9, 11, and 13 $\sigma$ levels. The vertical line marks the reference point. The horizontal line indicates the systemic velocity.}
\end{figure}

To better illustrate the gas kinematics and investigate whether the velocity gradient is consistent with rotation, we checked the PV cuts in the direction of the velocity gradient (see Fig \ref{fig:pvmap}). The plots of four identified disk candidates (Field3 MM1, Field8 MM1, Field18 MM1, and Field21 MM1) roughly exhibit a butterfly-shaped feature that is characteristic of Keplerian rotation. We overlaid red Keplerian curves on the PV plots of these four sources. It should be noted that these Keplerian curves are shown only for comparison purposes and are not fitting models. Precise mass estimates of the central protostars from the P-V plots still need higher-resolution observations to resolve the “true disks,” and to constrain the inclination of the disk plane. The rest plots show features that more closely resemble rigid-body rotation than Keplerian rotation. These features are also seen in simulations \citep[e.g.,][]{2019A&A...632A..50A} and observational data \citep[e.g.,][]{2017A&A...602A..59C} with comparable resolutions to our SMA observations. This is likely caused by the insufficient resolution. The contributions of unresolved disks and infalling, rotating envelopes are heavily blended, and thus make the Keplerian curve not fit well with the observational data. Higher-resolution observations are needed to separate the contributions of the disk and envelope.

\section{Discussion}\label{sec:discussion}
In recent years, many disk candidates around massive protostars have been detected \citep{2016A&ARv..24....6B}. The resolution of these observations is typically on the order of 1000 au or coarser, comparable to our SMA observations. However, it is worth noting that these disk-like structures are more likely infalling, rotating envelopes or “toroids” \citep{2007prpl.conf..197C, 2016A&ARv..24....6B}, rather than true accretion disks on a smaller scale. The true accretion disks should generally have sizes on the order of 100 au, as revealed by recent ALMA observations \citep[e.g.,][]{2018A&A...620A..31M, 2018ApJ...869L..24I, 2018ApJ...860..119G, 2020ApJ...897L..33J}. Since circumstellar disks are usually believed to mediate material onto protostars from the rotating envelopes, the detection of the rotating structures suggests the existence of embedded circumstellar disks.

With high-resolution SMA observations ($\sim1.8\arcsec$), we built up a unique sample consisting of 49 dust condensations associated with molecular outflows. Considering the well-established disk-outflow association in various environments \citep{2007prpl.conf..197C}, we expected to find evidence of a rotating disk or envelope in each outflow-associated condensation. However, only 27 condensations were detected in high-density tracers, and these dense-gas-traced condensations are our targets in searching for rotating evidence. Among the dense-gas-traced condensations, nine condensations show velocity gradients perpendicular to the outflow axis, indicating that these sources may host rotating envelopes or toroids with circumstellar disks embedded in them. The properties of these identified disk candidates are discussed in \ref{subsec:diskcand}. The other dense-gas-traced condensations have no clear velocity gradients or show velocity gradients parallel to the CO outflow axis, which is against the disk hypothesis. The missing rotation signs in these sources are discussed in \ref{subsec:non-rotation}. For the rest of the outflow-associated condensations, we do not detect appropriate molecular line emission for studying the kinematics of dense gas around these sources. More sensitive observations are needed to confirm whether these sources are disk candidates.

\subsection{Properties of identified disk candidates}\label{subsec:diskcand}

\citet{2021ApJ...918L...4C} determined the dust temperature of identified dust condensations (refer to their Appendix B) through a combination of two results: the large-scale ($\sim36\arcsec$) dust temperature map from SED fitting results of \emph{Herschel} data and a power law radial temperature profile $T(r)\propto r^{-0.4}L_*^{0.2}$ obtained from an approximate radiative transfer model, where $L_*$ represents the luminosity of the central heating source (protostars). The results are listed in Table \ref{tab:outflowassociated}. It is important to note that since the \emph{Herschel} data cannot resolve the MDCs, the derived dust temperature can be used as a lower limit when a central heating source is lacking. Additionally, $\mathrm{CH_3CN}$ and $\mathrm{H_2CO}$ lines have been detected in certain sources, and we employed the XCLASS (eXtend CASA Line Analysis Software Suite tool, \citet{2017A&A...598A...7M}) to estimate the molecular gas temperature of these sources. Detailed information on the molecular gas temperature calculation can be found in Appendix \ref{appe:temp}.

For the gas mass estimates, \citet{2021ApJ...918L...4C} assumed optically thin dust emission and adopted the equation:
\begin{equation}
\centering
M_{\mathrm{gas}}=\frac{gF_\nu D^2}{B_\nu(T)\kappa_\nu},
\end{equation}
where $M_\mathrm{gas}$ is the gas mass of the condensation, g is the gas-to-dust mass ratio,  $F_\nu$ the dust flux density at frequency, $\nu$, D the source distance that is set to 1.4 kpc, $B_\nu(T)$ the Planck function at dust temperature, \emph{T}, and $\kappa_\nu$ the dust opacity coefficient. They adopted a gas-to-dust mass ratio of 100 and a dust opacity, $\kappa_\nu=10(\nu/1\mathrm{THz})^\beta$ in $\mathrm{cm^2\ g^{-1}} $, following the HOBYS consortium\footnote{The Herschel imaging survey of OB young stellar objects. \url{http://www.herschel.fr/cea/hobys/en/index.php}} \citep[e.g.,][]{2017A&A...602A..77T}. The dust opacity power law index $\beta$ used here is two. The gas mass of disk candidates ranges from $4M_\odot$ to $80M_\odot$.

\subsection{Why don't all outflow sources show evidence of rotation?}\label{subsec:non-rotation}
Given that accretion disks are physically connected to outflows, we expected to detect evidence of rotating disks in all outflow-associated condensations. For sources without the emission of dense gas tracers, we cannot determine their kinematics. These sources were excluded from our analyses. So, our analysis focused on the 27 dense-gas-traced condensations. The disk candidate detection rate in the dense-gas-traced group is only 9/27=33.3\%, and the question is why only one third of outflow sources show evidence of disks.

\clearpage
\onecolumn
\begin{landscape}
\begin{longtable}{lccccllllllc}
\caption{\label{tab:outflowassociated}}\\
\multicolumn{12}{c}{Information on the dust condensations associated with outflow in our SMA observations}\\
\hline
\hline
Condensation & R.A. & Dec & ${T_\mathrm{dust}}$ \tablefootmark{a} & ${M_\mathrm{gas}}$\tablefootmark{b} & Size & $\mathrm{PA}_\mathrm{grad}$\tablefootmark{c} & $\mathrm{PA}_\mathrm{out}$ & Molecule\tablefootmark{d} & Radio\tablefootmark{e} & Class\tablefootmark{f} & Group\tablefootmark{g} \\
& (J2000) & (J2000) & (K) & $(M_\odot)$ & (pc) & (deg) & (deg) & Species & Source &  &  \\
\hline 
\endfirsthead

\multicolumn{12}{c}{(Continuued)}\\
\hline
\hline
Condensation & R.A. & Dec & ${T_\mathrm{dust}}$ \tablefootmark{a} & ${M_\mathrm{gas}}$\tablefootmark{b} & Size & $\mathrm{PA}_\mathrm{grad}$\tablefootmark{c} & $\mathrm{PA}_\mathrm{out}$ & Molecule\tablefootmark{d} & Radio\tablefootmark{e} & Class\tablefootmark{f} & Group\tablefootmark{g} \\
& (J2000) & (J2000) & (K) & $(M_\odot)$ & (pc) & (deg) & (deg) & Species & Source &  &  \\
\hline 
\endhead

\hline
\endfoot

\hline
\hline
\endlastfoot
Field3 MM1 & 20:36:52.17 & +41:36:24.09 & 40 & 6.6 & 0.016 & 28 & 132 & DCN/$\mathrm{H_2CO}$ & UC/HC & Class I\tablefootmark{*}  & Disk Candidate \\
Field4 MM1 & 20:36:57.67 & +42:11:30.18 & 37 & 40 & 0.019 & 40 & 127 & DCN/$\mathrm{H_2CO}$/$\mathrm{CH_3OH}$/ & Jet/Wind & … & Disk Candidate \\
 &  &  &  &  &  &  & & $\mathrm{CH_3CN}$ &  &  & \\
Field5 MM1 & 20:37:00.96 & +41:34:55.79 & 31 & 6.5 & 0.016 & 156 & 75 & DCN/$\mathrm{H_2CO}$/$\mathrm{CH_3OH}$/ & Jet/Wind & Class I\tablefootmark{*}  & Disk Candidate \\
 &  &  &  &  &  &  & & $\mathrm{CH_3CN}$ &  &  & \\
Field7 MM2 & 20:38:36.49 & +42:37:33.79 & 35 & 10.9 & 0.016 & 140 & 65 & DCN/$\mathrm{H_2CO}$/$\mathrm{CH_3OH}$/ & UC/HC & Class I\tablefootmark{*} & Disk Candidate \\
 &  &  &  &  &  &  & & $\mathrm{CH_3CN}$ &  &  & \\
Field8 MM1 & 20:39:16.74 & +42:16:09.31 & 40 & 9.1 & 0.015 & 1 & 80 & $\mathrm{C^{18}O}$/DCN/$\mathrm{H_2CO}$ & Jet/Wind & FS & Disk Candidate \\
Field9 MM1 & 20:40:05.39 & +41:32:13.02 & 17 & 84.2 & 0.011 & 130 & 43 & DCN/$\mathrm{H_2CO}$/$\mathrm{CH_3OH}$/ & Jet/Wind & Class I & Disk Candidate \\
 &  &  &  &  &  &  & & $\mathrm{CH_3CN}$ &  &  & \\
Field16 MM1 & 20:24:31.68 & +42:04:22.51 & 32 & 15.2 & 0.016 & 16 & 120 & DCN/$\mathrm{H_2CO}$/$\mathrm{CH_3OH}$/ & Jet/Wind & Class I\tablefootmark{*} & Disk Candidate \\
 &  &  &  &  &  &  & & $\mathrm{CH_3CN}$ &  &  & \\
Field18 MM1 & 20:20:44.65 & +39:35.20.11 & 27 & 3.7 & 0.021 & 126 & 30 & DCN/$\mathrm{H_2CO}$ & Jet/Wind & Class I & Disk Candidate \\
Field21 MM1 & 20:31:12.90 & +40:03:22.77 & 34 & 15.2 & 0.022 & 136 & 60 & DCN/$\mathrm{H_2CO}$/$\mathrm{CH_3OH}$ & Jet/Wind & Class I & Disk Candidate \\
Field1 MM1 & 20:35:34.62 & +42:20:08.95 & 18 & 21.4 & 0.019 & … & 130 & DCN/$\mathrm{H_2CO}$/$\mathrm{CH_3OH}$ & N & … & Dense-Gas-Traced \\
Field1 MM2 & 20:35:34.41 & +42:20:06.93 & 25 & 7.6 & 0.015 & 16 & 175 & DCN/$\mathrm{H_2CO}$/$\mathrm{CH_3OH}$ & N & Class I & Dense-Gas-Traced \\
Field4 MM3 & 20:36:57.16 & +42:11:32.73 & 18 & 0.9 & 0.013 & 28 & 22 & DCN/$\mathrm{H_2CO}$/$\mathrm{CH_3OH}$ & N & … & Dense-Gas-Traced \\
Field10 MM3 & 20:40:28.51 & +41:57:15.86 & 18 & 3.4 & 0.015 & … & 150 & $\mathrm{CH_3OH}$ & N & … & Dense-Gas-Traced \\
Field11 MM1 & 20:40:33.57 & +41:59:01.07 & 37 & 4.2 & 0.016 & ... & 73 & DCN/$\mathrm{H_2CO}$/$\mathrm{CH_3OH}$ & Jet/Wind & Class I & Dense-Gas-Traced \\
Field12 MM1 & 20:40:33.93 & +41:51:03.99 & 22 & 1.9 & 0.014 & … & 123 & $\mathrm{H_2CO}$ & Jet/Wind & Class I\tablefootmark{*} & Dense-Gas-Traced \\
Field14 MM1 & 20:20:30.63 & +41:21:26.27 & 42 & 13.1 & 0.015 & 87\tablefootmark{g} & 175 & DCN/$\mathrm{H_2CO}$/$\mathrm{CH_3OH}$ & UC/HC & Class I\tablefootmark{*} & Dense-Gas-Traced \\
Field17 MM3 & 20:20:39.36 & +39:38:11.76 & 21 & 8.5 & 0.019 & 25 & 33 & $\mathrm{H_2CO}$ & Jet/Wind & … & Dense-Gas-Traced \\
Field22 MM2 & 20:31:20.67 & +38:57:23.46 & 18 & 10.3 & 0.023 & 30 & 34 & $\mathrm{H_2CO}$/$\mathrm{CH_3OH}$ & Jet/Wind & … & Dense-Gas-Traced \\
Field23 MM1 & 20:31:58.15 & +40:18:36.19 & 26 & 1.3 & 0.014 & … & 29 & $\mathrm{H_2CO}$ & Jet/Wind & Class I & Dense-Gas-Traced \\
Field26 MM1 & 20:30:29.13 & +41:15:53.93 & 21 & 10.4 & 0.019 & … & 10 & $\mathrm{H_2CO}$/$\mathrm{CH_3OH}$ & Jet/Wind & … & Dense-Gas-Traced \\
Field29 MM2 & 20:28:09.60 & +40:52:51.00 & 23 & 2.6 & 0.019 & 116 & 126 & $\mathrm{H_2CO}$/$\mathrm{CH_3OH}$ & N & … & Dense-Gas-Traced \\
Field32 MM1 & 20:32:28.56 & +40:19:41.52 & 43 & 1.6 & 0.010 & ... & 10 & $\mathrm{H_2CO}$ & N & FS & Dense-Gas-Traced \\
Field32 MM5 & 20:32:21.46 & +40:20:14.51 & 16 & 8.3 & 0.014 & 87 & 105 & $\mathrm{H_2CO}$/$\mathrm{CH_3OH}$ & Jet/Wind & … & Dense-Gas-Traced \\
Field33 MM2 & 20:39:03.00 & +42:25:51.04 & 37 & 32.0 & 0.020 & … & -160 & DCN/$\mathrm{H_2CO}$/$\mathrm{CH_3OH}$ & N & … & Dense-Gas-Traced \\
Field33 MM3 & 20:39:03.23 & +42:25:51.37 & 18 & 20.0 & 0.019 & … & 10 & DCN/$\mathrm{H_2CO}$/$\mathrm{CH_3OH}$  & N & … & Dense-Gas-Traced \\
Field33 MM16 & 20:39:01.03 & +42:22:48.72 & 35 & 14.0 & 0.017 & ... & 110 & DCN/$\mathrm{H_2CO}$/$\mathrm{CH_3OH}$ & UC/HC & Class I & Dense-Gas-Traced \\
Field33 MM17 & 20:39:00.43 & +42:22:46.56 & 27 & 43.0 & 0.029 & 70 & 60 & DCN/$\mathrm{H_2CO}$/$\mathrm{CH_3OH}$ & N & … & Dense-Gas-Traced \\
Field2 MM4 & 20:36:07.55 & +41:40:08.56 & 42 & 0.3 & 0.017 & … & 90 & ... & Jet/Wind & Class I\tablefootmark{*} & Outflow-Associated \\
Field3 MM2 & 20:36:52.20 & +41:36:08.80 & 24 & 5.5 & 0.015 & … & 72 & ... & Jet/Wind & … & Outflow-Associated \\
Field6 MM2 & 20:38:03.27 & +42:40:03.64 & 16 & 1.7 & 0.014 & … & 157 & ... & Jet/Wind & … & Outflow-Associated \\
Field8 MM3 & 20:39:17.87 & 42:16:10.10 & 17 & 0.6 & 0.015 & … & 40 & ... & N & … & Outflow-Associated \\
Field10 MM2 & 20:40:28.56 & +41:57:11.23 & 23 & 1.4 & 0.015 & … & 80 & ... & N & Class I & Outflow-Associated \\
Field10 MM4 & 20:40:28.96 & +41:57:07.90 & 26 & 0.8 & 0.017 & … & 80 & ... & N & Class I & Outflow-Associated \\
Field11 MM3 & 20:40:33.72 & +41:58:53.81 & 23 & 1.2 & 0.015 & … & 150 & ... & N & … & Outflow-Associated \\
Field14 MM2 & 20:20:30.12 & +41:22:07.00 & 18 & 4.3 & 0.017 & … & 130 & ... & N & … & Outflow-Associated \\
Field17 MM4 & 20:20:37.93 & +39:38:20.86 & 19 & 10.3 & 0.020 & … & 10 & ... & N & … & Outflow-Associated \\
Field17 MM5 & 20:20:39.35 & +39:37:52.43 & 22 & 7.2 & 0.026 & … & 30 & ... & Jet/Wind & … & Outflow-Associated \\
Field17 MM8 & 20:20:38.90 & +39:38:18.62 & 42 & 2.6 & 0.029 & … & 120 & ... & N & Class I & Outflow-Associated \\
Field18 MM3 & 20:20:44.89 & +39:35:26.39 & 20 & 0.5 & 0.018 & … & 130 & ... & N & … & Outflow-Associated \\
Field20 MM2 & 20:29:57.67 & +40:16:04.22 & 22 & 0.9 & 0.017 & … & 150 & ... & N & Class II & Outflow-Associated \\
Field21 MM3 & 20:31:12.42 & +40:03:10.24 & 30 & 1.0 & 0.020 & … & 95 & ... & N & Class I\tablefootmark{*} & Outflow-Associated \\
Field21 MM4 & 20:31:14.62 & +40:03:05.91 & 35 & 1.6 & 0.023 & … & -50 & ... & N & … & Outflow-Associated \\
Field21 MM7 & 20:31:13.37 & +40:03:10.39 & 38 & 1.5 & 0.026 & … & 40 & ... & Jet/Wind & … & Outflow-Associated \\
Field22 MM3 & 20:31:21.05 & +38:57:16.04 & 18 & 3.1 & 0.022 & … & -30 & ... & N & … & Outflow-Associated \\
Field25 MM2 & 20:32:41.00 & +38:46:36.44 & 15 & 4.9 & 0.032 & … & 150 & ... & N & … & Outflow-Associated \\
Field26 MM2 & 20:30:26.88 & +41:16:09.00 & 21 & 4.9 & 0.015 & … & 45 & ... & N & … & Outflow-Associated \\
Field28 MM1 & 20:32:21.05 & +41:07:54.12 & 30 & 2.2 & 0.014 & … & 150 & ... & Jet/Wind & Class I & Outflow-Associated \\
Field28 MM2 & 20:32:23.76 & +41:07:57.23 & 28 & 1.4 & 0.015 & … & 30 & ... & N & Class I & Outflow-Associated \\
Field31 MM1 & 20:34:59.24 & +41:34:47.50 & 21 & 2.8 & 0.023 & … & 8 & ... & N & … & Outflow-Associated \\
\hline
\end{longtable}
\tablefoot{\\
\tablefoottext{a}{Dust temperature estimated by \citet{2021ApJ...918L...4C}. The 1-$\sigma$ uncertainty of \emph{T} is $\sigma_T=4$ K.}\\
\tablefoottext{b}{Gas mass estimated by \citet{2021ApJ...918L...4C}. Monte-Carlo simulations show that the 1-$\sigma$ uncertainty of $M_\mathrm{gas}$ is $\sim0.26$ dex ($\sim64\%$).}\\
\tablefoottext{c}{Orientation of the velocity gradient revealed by the selected line emission.}\\
\tablefoottext{d}{Detection of dense gas tracers, $\mathrm{CH_3CN}$, $\mathrm{CH_3OH}$, and $\mathrm{H_2CO}$, in each dust condensation associated with outflow. Information on the transitions of these molecules is listed in Table \ref{tab:lineinfo}. "..." means no dense gas tracers were detected as showing the kinematics. "$\mathrm{C^{18}O}$" means the $\mathrm{C^{18}O}$ (2-1) emission in Field8 MM1 also traces dense gas surrounding the condensation and reveals the evidence of a rotating structure.}\\
\tablefoottext{e}{Radio sources identified by \citet{2022ApJ...927..185W}. “UC/HC” means the radio detection is likely from the ultra- or hyper-compact H II region. “Jet/Wind”
means the radio detection is likely from a radio jet or wind. “N” means no radio source coincides with the condensation.}\\
\tablefoottext{f}{IR Classification identified by \citet{2014AJ....148...11K}. "Class" implies the mid-infrared color of these condensations, which can in turn indicate the YSO evolutionary stage of these condensations. “FS” means a “flat spectrum” that has a spectral index, $\alpha$, ranging from -0.3 to 0.3. The discussions of the mid-infrared colors are shown in Section \ref{sec:evolutionstage}}\\
\tablefoottext{g}{Classifications of the condensations. “Outflow-associated” means these condensations are associated with outflow. “Dense-gas-traced” are condensations included in “outflow-associated” condensations that have emission of dense gas tracers. “Disk candidate” condensations are included in dense-gas-traced condensations showing evidence of a rotating disk. The classifications of the condensations are described in Section \ref{subsec:densegas}.}\\
\tablefoottext{g}{The clear velocity gradient from line emission away from a potential outflow central source. So, it is not identified as a disk candidate.}\\
\tablefoottext{*}{The sources are saturated or unresolved in a 24 $\mu\mathrm{m}$ image, which means photometry is unavailable.}
}
\end{landscape}
\twocolumn

\subsubsection{Non-detection of rotating structures}
There are several explanations for the lack of a velocity gradient toward some dense-gas-traced condensations. First, in contrast to the simplified classic paradigm of a stable disk-jet system, many simulations \citep{2013ApJ...766...97M,2015MNRAS.446.2776S,2016MNRAS.463.2553R} predict that protostellar cores may accrete material through asymmetrical filaments without a large, stable Keplerian disk, and some recent ALMA observations support this scenario \citep[e.g.,][]{2020ApJ...905...25G}. However, it needs higher-spatial-resolution observations to confirm whether these sources without rotating structures in our sample actually accrete material through filamentary flows. Second, several condensations, for example Field4 MM3 and Field10 MM3, are located in complex regions. The emission from adjacent energetic events (e.g., accretion flow or jets and outflows) may confuse the velocity field around these condensations and make it difficult to distinguish the rotating structures or toroids. On the other hand, relatively isolated sources, forming in an environment away from other energetic events, have a higher chance of showing detectable rotating structures \citep[e.g., Field9 MM1, Field16 MM1 in this work and G17.64+0.16 in][]{2018A&A...620A..31M}.

The inclination of the disk plane with respect to the line of sight may also affect the observational evidence of a rotating disk. We made a simple simulation following \citet{2019ApJ...873...73Z} to investigate the influence of inclination on disk identification. 

\begin{figure}[!h]
    \centering
    \includegraphics[width = 7cm]{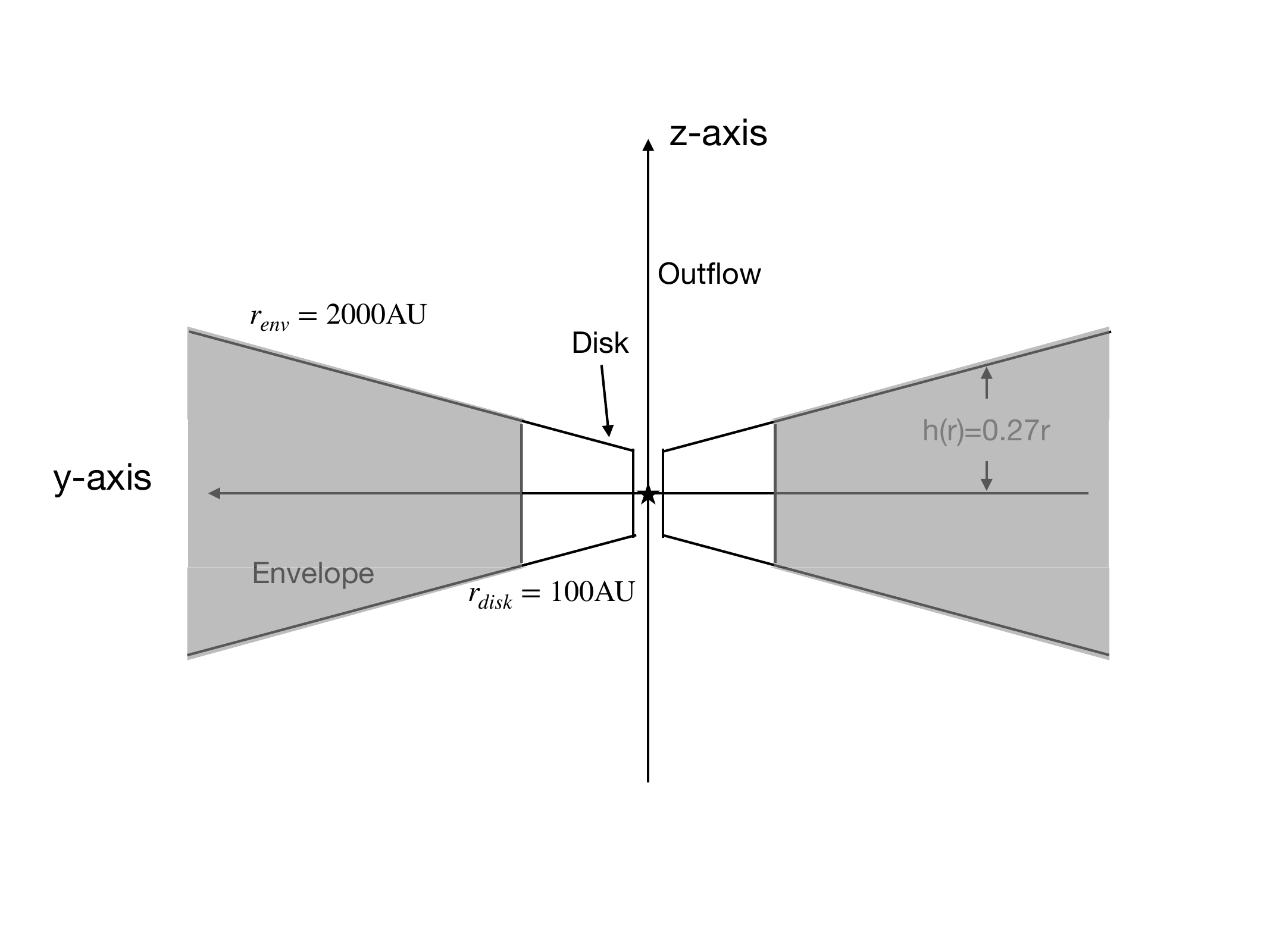}
    \caption{Schematic diagram of our envelope-disk model. The outflow axis is aligned with the z axis. The inner part of the Keplerian disk starts from the inner boundary, $r_\mathrm{in}=10$ au, and extends to the centrifugal barrier, $r_\mathrm{CB}=100$ au. The outer envelope ranges from the centrifugal barrier, $r_\mathrm{CB}$, to the outer boundary, $r_\mathrm{out}=2000$ au.}
    \label{fig:EnvDiskModel}
\end{figure}

\begin{figure}[!ht]
    \includegraphics[width=9cm]{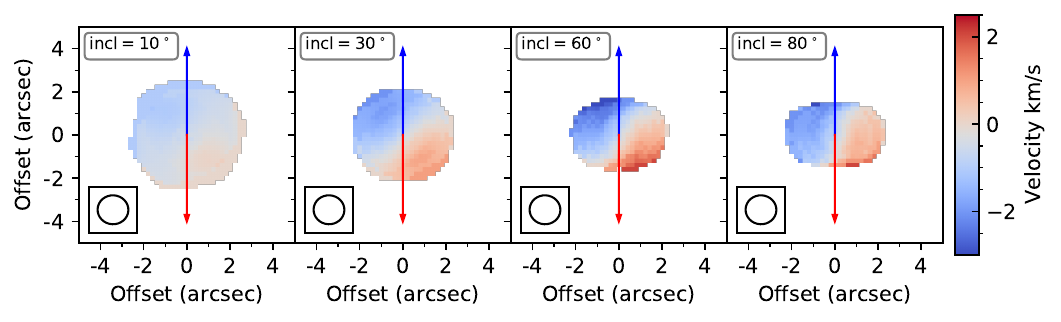}
    \caption{Synthetic SMA observations made with the CASA task “simobserve” from the outputs of radiative transfer calculations of the envelope-disk model shown in Fig \ref{fig:EnvDiskModel}. From left to right, the inclination angle of the rotation axis with respect to the line of sight changes from 10$^\circ$ to 80 $^\circ$. The color scale shows the $\mathrm{CH_3CN\ (12_3-11_3)}$ first-moment map. The red and blue arrows represent redshifted and blueshifted outflows, respectively. The bottom left of each panel shows the synthesized beam.}
    \label{fig:SimVel}
\end{figure}

Fig \ref{fig:EnvDiskModel} shows the schematic view of an inner Keplerian-rotating disk embedded in a larger-scale infalling and rotating envelope. The masses of the central protostar, accretion disk, and envelope are $10M_\odot$, $3M_\odot$, and $5M_\odot$, respectively. We assumed an opening angle of $150^\circ$ for the outflow cavity. The flattened envelope has a radial free-fall velocity. Its rotation velocity can be described as $v_\phi\propto r^{-1}$. The inner disk is assumed to be in Keplerian rotation ($v_\mathrm{rot}\propto r^{-1/2}$). We used open-source software, \emph{RADMC-3D} \citep{2012ascl.soft02015D}, and the CASA simulator task \emph{simobserve} to generate synthetic SMA observations for the model with different inclination angles. The settings for the synthetic observation are the same as for our SMA observations, with an angular resolution of $\sim$1.8" and a velocity resolution of $\mathrm{1\ km\ s^{-1}}$. Fig \ref{fig:SimVel} shows the first-moment map of $\mathrm{CH_3CN\ (12_3-11_3)}$ for the synthetic observation. The simulation indicates that only when the rotation axis is extremely inclined to the line of sight (incl $\leq10^\circ$) can the velocity structure of the disk and envelope not be revealed by our SMA observation, meaning the effect of the inclination angles is not the prominent cause of the non-detection of a rotating structure. However, it is worth noting that the orientation of the velocity gradient slightly changes with the inclination. When the disk plane is nearly edge-on (incl $\geq80^\circ$), the velocity gradient is almost perpendicular to the outflow axis. As the inclination angle decreases, the orientation of the velocity gradient slightly shifts toward the outflow axis. This phenomena is also seen in other simulations \citep[e.g.,][]{2019A&A...632A..50A}{}{}. We speculate that it may be because the envelope and disk contributions are strongly blended under poor resolutions. The velocity gradients seen at a low inclination angle (including $\leq30^\circ$) are mainly contributed by the infalling-rotating envelope since the line-of-sight velocity component in the disk is very small. The disk contributions become increasingly important as the structures get more inclined. Detailed kinematic analyses using high-resolution simulations and more sophisticated models are necessary to investigate the direction of the velocity gradient changes with the inclination and to what extent this could affect the detection of a rotating disk, which is beyond the scope of this work.

\subsubsection{Comparisons between disk candidates and others}\label{sec:evolutionstage}
Except for the examples mentioned above, we compared the properties of disk candidates and non-disk dense-gas-traced condensations to further explore the detectability of rotating structures around protostars.

\begin{figure}[!h]
    \includegraphics[width=9cm]{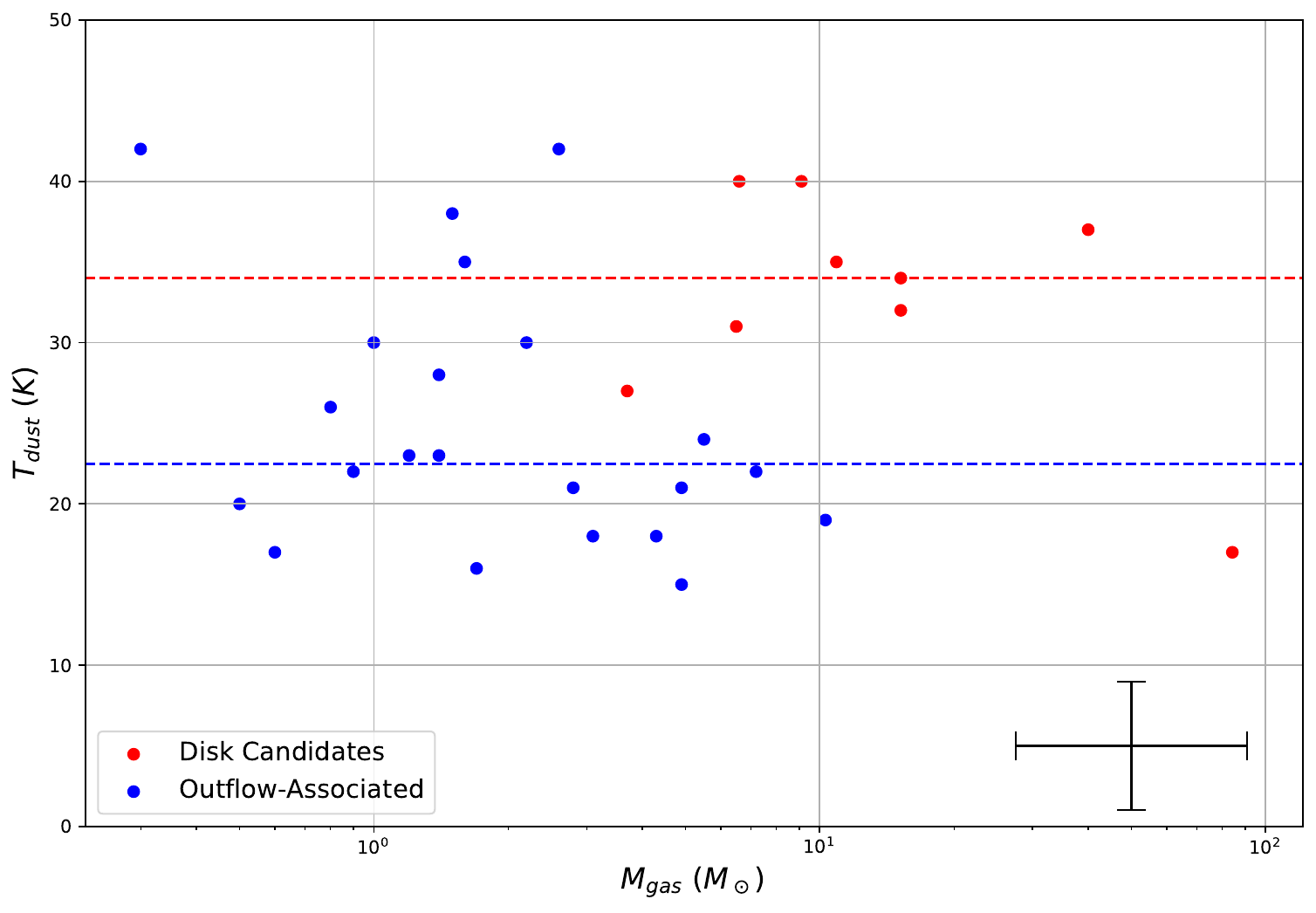}
    \caption{Dust temperature distribution of all dense-gas-traced condensations in relation of gas mass. Red and blue dots represent disk candidates and non-disk dense-gas-traced condensations, respectively. The dashed red line shows the median temperature of the disk candidates, 34 K. The dashed black line shows the median temperature of the non-disk dense-gas-traced condensations, 22.5 K. In the lower right corner, the vertical and horizontal black error bar indicates the uncertainties of the dust temperature ($\sigma_T=4$ K) and the gas mass ($\sim0.26$ dex) from \citet{2021ApJ...918L...4C}, respectively.}
    \label{fig:tdustdist}
\end{figure}

Table \ref{tab:outflowassociated} lists the parameters of the condensations. To better illustrate the temperature difference between disk and non-disk groups, Fig \ref{fig:tdustdist} shows the distribution of dust temperatures for all dense-gas-traced condensations. Most disk candidates, except for Field9 MM1 and Field18 MM1, have dust temperatures exceeding 30 K, with a median temperature of 34 K. On the other hand, the non-disk dense-gas-traced condensations generally have lower dust temperatures, with a median value of 22.5 K. The 1-$\sigma$ uncertainty of the dust temperature is $\sigma_T=4$ K. The difference between the median temperature of disk candidate and non-disk dense-gas-traced sources is about three times the temperature uncertainty. We ran a Kolmogorov-Smirnow test (K-S test) to investigate how significant the temperature difference is between the disk candidate and non-disk dense-gas-traced sources. A p value of $\approx0.01$ was obtained for testing the null hypothesis that the two distributions are identical, indicating a statistically significant difference in dust temperature between the two groups. Even though the dust temperature is derived from the \emph{Herschel} data with a coarser resolution, considering that the temperatures were uniformly derived for all the sources in our sample, we took it into account in checking the evolutionary stages of the sources. The increasing trend in the dust temperature from disk candidates to non-disk dense-gas-traced condensations suggests disk candidates are more evolved.

The brightness of radio continuum emission can also to some extent probe the evolution of high-mass star formation. \citet{2022ApJ...927..185W} has made a systematic study of the radio properties of the condensations in our sample with high-angular-resolution VLA observations. They find that the detection rate of radio sources, either in the form of radio jets or wind, or UC/HC \HII, increases as the protostellar object evolves. This trend is consistent with that in \citet{2016ApJS..227...25R}. It may come from this that, at a very early stage, the star formation activities are not strong enough to produce detectable radio emission. For example, \citet{2016ApJ...818...52T} modeled the evolution of a massive protostar and its associated jet as it is being photoionized by the protostar. In this outflow-confined \HII\ region model, at a very early stage, the ionizing photon luminosity is too low to ionize the outflow, and thus does not produce free-free emission. As a result, the sources without detectable radio emission are more likely at an earlier stage than the sources with radio emission.  We cross-checked 49 outflow-associated condensations with their identified radio sources and found that the radio detection rate of disk candidates is 100$\%$, substantially higher than the detection rate of non-disk dense-gas-traced sources, which stands at 9/18=50$\%$. Consequently, disk candidates are generally more evolved than non-disk dense-gas-traced sources, which is consistent with the speculation from the above comparison of dust temperatures.

To further check the evolutionary stages of the sources, we conducted a cross-correlation analysis between the 49 outflow-associated condensations and the mid-infrared sources from the Cygnus-X Archive catalog\footnote{\url{https://irsa.ipac.caltech.edu/cgi-bin/Gator/nph-dd}}. By applying a spatial constraint of 2'', approximately the mean size of the condensation, we found that 23 condensations coincide with \emph{Spitzer} sources (see Table \ref{tab:irinfo}). In a previous study, \citet{2014AJ....148...11K} systemically surveyed protostars in the Cygnus-X region using \emph{Spitzer} IRAC and 24 $\mu m$ photometry. They estimated the spectral index, $\alpha = \delta\log(\lambda F_\lambda)/\delta\log(\lambda)$ in wavelength ranges from $3.6\ \mu m$ to $24\ \mu m$. Based on their estimated spectral index, $\alpha$, these sources are identified as Class 0/I ($\alpha > 0.3$), flat spectrum ($-0.3<\alpha<0.3$), and Class II \citep[using criteria of][]{2009ApJS..184...18G} young stellar objects (YSOs). We note that \citet{2014AJ....148...11K} classified these protostar candidates only depending on their mid-infrared color, without considering their masses. Therefore, it is important to clarify that a condensation classified as Class 0/I/II or flat spectrum only reflects its IR color and does not mean that it is a low-mass protostar or young star. We used this information to constrain the evolutionary stages of these sources. Apparently, a source dark all the way from 3.6 um to 24 um is cold and at an earlier evolutionary stage compared to a source that is detected in several Spitzer bands and that could be classified as Class 0/I/II or flat spectrum by \citet{2014AJ....148...11K}. Some of the brightest sources in our sample are saturated (e.g., Field7 MM2) or unresolved (e.g., Field2 MM2) at $24\ \mu \mathrm{m}$, and thus it is not possible to do 24 $\mu$m photometry. \citet{2014AJ....148...11K} excluded these sources in their classification catalog, while we notice that these sources always have bright emission at $3.6\ \mu \mathrm{m\ to\ } 8.0\ \mu \mathrm{m}$. Therefore, it is reasonable to assume that the sources that are saturated or unresolved at $24\ \mu \mathrm{m}$ are also relatively more evolved. Eventually, we found that more than half of the non-disk (11/18=61.1$\%$) dense-gas-traced condensations are dark in 3.6 $\mathrm{\mu m}$ to 24 $\mathrm{\mu m}$, while 8/9=88.9$\%$ of the disk candidates have bright emission in 3.6 $\mathrm{\mu m}$ to 24 $\mathrm{\mu m}$. This again suggests that disk candidates are relatively more evolved than non-disk dense-gas-traced sources.

\begin{table*}[!ht]
\centering
\caption{Infrared properties of outflow-associated condensations}
\resizebox{\textwidth}{!}{
\begin{tabular}{lllllllll}
\hline
\hline
Condensation & R.A. & Dec & $3.6\ \mu m$ & $4.5\ \mu m$ & $5.8\ \mu m$ &$8.0\ \mu m$ & $24\ \mu m$ & Class\tablefootmark{a}\\
 & (deg) & (deg) & (mag) & (mag) & (mag) & (mag) & (mag) & \\
\hline
Field1 MM2 & 308.893524 & 42.335224 & $11.00\pm 0.03$ & $8.16\pm0.02 $ & $6.81\pm 0.02$ & $6.09\pm 0.03 $ & $1.55\pm 0.07 $ & Class I \\
Field2 MM4 & 309.031372 & 41.669178 & $5.23\pm0.02$ & $4.29\pm 0.02$ & $3.16\pm 0.02$ & $2.66\pm 0.02 $ & ...\tablefootmark{*} & ... \\
Field3 MM1 & 309.217438 & 41.606693 & $7.37\pm 0.02$ & $5.46\pm0.02 $ & $4.016\pm 0.02 $ & $2.677\pm 0.02 $ & ...\tablefootmark{*} & ... \\
Field5 MM1 & 309.253937 & 41.582123 & $8.92\pm 0.03$ & $6.69\pm 0.02$ & $5.13\pm 0.02 $ & $4.19\pm 0.05$ & ...\tablefootmark{*}  & ... \\
Field7 MM2 & 309.652039 & 42.625977 & ... & $6.24\pm 0.03$ & $4.55 \pm 0.04 $ & $3.55\pm 0.04$ & ...\tablefootmark{*} & ... \\
Field8 MM1 & 309.819672 & 42.269257 & $6.49\pm 0.02$ & $5.17\pm 0.02$ & $4.00 \pm 0.02$ & $3.24\pm 0.02 $ & $0.255\pm 0.02 $ & FS \\
Field9 MM1 & 310.02247 & 41.53695 & ... & $12.07\pm 0.07$ & $11.36 \pm 0.11$ & $11.35\pm 0.19 $ & $3.02\pm 0.02 $ & Class I \\
Field10 MM2 & 310.11898 & 41.95312 & $10.83\pm0.03$ & $9.01\pm 0.03$ & $8.27\pm 0.03 $& $7.95\pm 0.02 $ & $1.04\pm 0.02$ & Class I \\
Field10 MM4 & 310.120758 & 41.952198 & $8.79\pm0.02$ & $7.38\pm 0.02$ & $6.37\pm 0.02 $& $5.44\pm 0.02 $ & $1.19\pm 0.03$ & Class I \\
Field11 MM1 & 310.139526 & 41.983509 & $14.38\pm 0.05$ & $11.52\pm 0.02$ & $11.68\pm 0.08 $& ... & $3.91\pm 0.1 $ & Class I \\
Field12 MM1 & 310.141357 & 41.851223 & $10.28\pm0.02$ & $8.71 \pm 0.02$ & $7.32\pm 0.02 $ & $6.32\pm 0.03 $ & ...\tablefootmark{*}  & ... \\
Field14 MM1 & 305.127655 & 41.357201 & $5.16\pm 0.02$ & $4.34\pm 0.02 $ & $2.20\pm 0.02 $ & ...\tablefootmark{*} & ...\tablefootmark{*}  & ... \\
Field16 MM1 & 306.132111 & 42.072918 & $10.80\pm 0.03$ & $7.29\pm0.02 $ & $5.24\pm 0.02 $ & $3.97\pm 0.01 $ & ...\tablefootmark{*}  & ... \\
Field17 MM8 & 305.162201 & 39.638287 & $12.07\pm0.12$ & $10.08\pm 0.04$ & $8.43\pm 0.04 $ & $7.03\pm 0.04 $ & $1.43\pm 0.08 $ & Class I \\
Field18 MM1 & 305.185699 & 39.589119 & $9.09\pm 0.02$ & $7.80\pm 0.02$ & $6.82\pm 0.04 $ & $5.87\pm 0.06 $ & $2.02\pm 0.1 $ & Class I \\
Field20 MM2 & 307.490326 & 40.267841 & $8.45\pm0.02$ & $7.61\pm 0.02$ & $6.99\pm 0.02 $ & $6.26\pm 0.02 $ & $3.39\pm 0.04 $ & Class II \\
Field21 MM1 & 307.803802 & 40.056416 & ... & ... & ... & $8.48\pm 0.08$ & $1.343\pm 0.07 $ & Class I \\
Field21 MM3 & 307.801636 & 40.052952 & $10.76\pm0.04$ & $9.03\pm 0.02 $ & $7.18\pm 0.04 $ & $5.80\pm 0.08 $ & ...\tablefootmark{*}  & ... \\
Field23 MM1 & 307.99231 & 40.310028 & $13.98\pm0.03$ & $12.19\pm 0.03$ & $11.35\pm 0.04$ & $10.97\pm 0.11 $ & $3.68\pm 0.04 $ &  Class I \\
Field28 MM1 & 308.087738 & 41.131798 & $11.12\pm0.02$ & $8.53\pm 0.02 $ & $6.48\pm 0.02 $ & $5.07\pm 0.02 $ & $0.87\pm 0.02 $ &  Class I \\
Field28 MM2 & 308.098999 & 41.132629 & $13.63\pm0.04$ & $10.82\pm 0.02 $ & $9.03\pm 0.02 $ & $8.01\pm 0.04 $ & $4.83\pm 0.1 $  & Class I \\
Field32 MM1 & 308.119202 & 40.328213 & $7.37\pm0.02$ & $6.30\pm 0.02$ & $5.35\pm 0.02$ & $4.43\pm 0.02 $ & $0.98\pm 0.02$ & FS \\
Field33 MM16 & 309.754211 & 42.380585 & $11.10\pm 0.04$ & $10.10\pm 0.05 $ & $7.48\pm 0.04 $ & $5.67\pm 0.05 $ & $-0.24\pm0.03$  & Class I \\
\hline
\end{tabular}
}
\tablefoot{\\
Infrared properties from Cygnus-X Archive catalog.\\
\tablefoottext{a}{Classification results from \citet{2014AJ....148...11K}. “FS” means flat spectrum. “...” means that these IR sources were not selected by \citet{2014AJ....148...11K} since the 24 $\mathrm{\mu m}$ photometry is unavailable for these sources.}\\
\tablefoottext{*}{“*” means the emission is saturated or unresolved in this band.}
}
\label{tab:irinfo}
\end{table*}

Several studies \citep[e.g.,][]{2006ApJ...646.1070A,2019ApJ...871..141Q,2021A&A...648A..41V,2022ApJ...927...88H} have found that the outflow opening angles increase as the outflow central sources evolve. Within our sample, we notice that the outflows emanating from disk candidates tend to have wide-angle morphology (e.g., Field8 MM1) or irregular shapes (e.g., the blueshifted lobe associated with Field7 MM2), while the outflows originating in non-disk sources always appear to be collimated (e.g., Field1 MM1). This provides another piece of evidence that disk candidates are generally more evolved.

From all the perspectives discussed above, dust condensations with rotating disks or toroids are generally more evolved than other sources. The apparent lack of rotating structures at relatively early evolutionary stages might be due to the outflow-envelope interactions proposed by \citet{2006ApJ...646.1070A}. Table \ref{tab:outflowassociated} shows that some non-disk dense-gas-traced condensations display velocity gradients rougly parallel to the axis of outflow (e.g., Field1 MM2). For those dense-gas-traced condensations without a clear velocity gradient, their dense gas emission always appear to be elongated along the outflow axis. Since the spatial resolution of our SMA observations is insufficient to resolve the true accretion disk (on the order of $10^2\ \mathrm{au}$) that is embedded within the envelope, the high-density tracers mainly come from the contribution of the high-density envelope (on the order of $10^3\ \mathrm{au}$). In young, deeply embedded sources, the powerful outflow entrains and pushes gas into the envelope, causing the elongated distribution of the dense gas along the outflow axis. As the protostar evolves, the outflow cavities tend to widen. Most of the remaining circumstellar gas is concentrated in the flattened envelope perpendicular to the outflow axis, which is consistent with the model from \citet{2019ApJ...873...73Z}. At this stage, it is more common to detect rotating structures or toroids with dense gas tracers.

\section{Summary}\label{sec:summary}
We have shown the results of SMA observations of continuum and molecular line emission toward 31 single-pointing fields and two mosaic fields, covering 48 MDCs in Cygnus-X. The high-resolution observations resolve these MDCs into about 200 dust condensations with sizes of order 0.01 pc, 49 of which are found to be associated with CO outflows. The velocity fields for 27 of these outflow-associated condensations are revealed by high-density tracers (e.g., $\mathrm{CH_3CN}$ and $\mathrm{CH_3OH}$ lines). Eventually, nine dust condensations are recognized to be rotating, as is evidenced by them showing a velocity gradient perpendicular to the outflow axis. The detected rotating condensations are more likely “toroids” or envelopes with radii on the order of 1000 au, enclosing smaller disks. Further investigations of the P-V diagrams along the velocity gradient reveal cases with butterfly-shaped patterns, indicative of Keplerian-like rotation.

We generated synthetic SMA observations of a disk-envelope model at varying inclination angles with \emph{RADMC-3D}. Even though the model is simplified, the simulation suggests that inclination can hinder the detection of rotating envelopes or toroids only when the rotation axis is largely inclined to the line of sight.

By comparing disk candidates and non-disk dense-gas-traced condensations, we conclude that the disk candidate detection rate could be sensitive to protostellar evolution. In young, deeply embedded sources, the powerful outflows may dominate the kinematics of the envelopes. As the protostar evolves, the outflows carve a wide-angle cavity, resulting in a flattened rotating envelope that is easier to detect. The insufficient angular resolution of our SMA observations makes it difficult to reveal the deeply embedded “true disk” that has a small size. Observations with a higher resolution and sensitivity are required to figure out whether these disk candidates actually host small-scale Keplerian-like disks and the reason why other dense-gas-traced and outflow-associated condensations do not show evidence of rotating structures.
\begin{acknowledgements}
    This work is supported by National Key R\&D Program of China No. 2022YFA1603100, No. 2017YFA0402604, the National Natural Science Foundation of China (NSFC) grant U1731237, and the science research grant from the China Manned Space Project with No. CMS-CSST-2021-B06.
\end{acknowledgements}


\bibliography{xingpandraft}{}
\bibliographystyle{aa} %


\begin{appendix}
\onecolumn
\section{Dust continuum and CO outflow maps}\label{appe:dustcont}
\begin{table*}[!ht]
\renewcommand{\arraystretch}{1.1}
\centering
\caption{Physical parameters of the MDCs}
\label{tab:mdcpara}
{
\begin{tabular}{lllllc}
\hline
\hline
Field & Motte07\tablefootmark{b}& MDC\tablefootmark{c} & R.A.  & Decl. \\
& & & (J2000) & (J2000)\\
\hline 
1 & N02/N03 & 220 & 20:35:34.630 & +42:20:08.787 \\
2 & N05/N06 & 274 & 20:36:07.300 & +41:39:57.994 \\
3 & N10 & 725 & 20:36:52.199 & +41:36:22.991 \\
4 & N12/N13 & 248 & 20:36:57.397 & +42:11:27.997 \\
5 & N14 & 714 & 20:37:00.900 & +41:34:57.002 \\
6 & N22/N24 & 1267 & 20:38:04.599 & +42:39:53.997 \\
7 & N30/N31/N32 & 1112/1231 & 20:38:37.401 & +42:37:32.994 \\
8 & N56 & 698/1179 & 20:39:16.897 & +42:16.07.003 \\
9 & N63 & 341 & 20:40:05.202 & +41:32:13.003 \\
10 & N64/N65 & 801 & 20:40:28.397 & +41:57:10.997 \\
11 & N68 & 684 & 20:40:33.499 & +41:59:02.995 \\
12 & N69 & 4315/4797 & 20:40:33.698 & +41:50:59.000 \\
13 & NW01/NW02 & 327/742 & 20:19:38.998 & +40:56:45.006 \\
14 & NW04/NW05/NW07 & 640/675 & 20:20:30.503 & +41:21:39.998 \\
15 & NW12 & 839 & 20:24:14.301 & +42:11:43.001 \\
16 & NW14 & 310 & 20:24:31.701 & +42:04:22.999 \\
17 & S06/S07/S08/S09 & 507/753 & 20:20:38.599 & +39:38:00.006 \\
18 & S10 & 798 & 20:20:44.400 & +39:25:19.999 \\
19 & S15 & 874 & 20:27:13.996 & +37:22:57.997 \\
20 & S29 & 723 & 20:29:58.297 & +40:15:57.994 \\
21 & S30/S31 & 509 & 20:31:12.599 & +40:03:15.996 \\
22 & S32 & 351 & 20:31:20.303 & +38:57:16.002 \\
23 & S34 & 1225 & 20:31:57.801 & +40:18:30.004 \\
24 & S41 & 892 & 20:32:33.397 & +40:16:42.997 \\
25 & S42/S43 & 540 & 20:32:40.799 & +38:46:30.994 \\
26 & ... & 214/247 & 20:30:28.503 & +41:15:55.000 \\
27 & ... & 302/520 & 20:35:09.499 & +41:13:29.995 \\
28 & ... & 340 & 20:32:22.500 & +41:07:55.996 \\
29 & ... & 370 & 20:28:09.402 & +40:52:49.995 \\
30 & ... & 608 & 20:34:00.003 & +41:22:25.001 \\
31 & ... & 1460/2320 & 20:35:00.002 & +41:34:57.002 \\ 
\hline
32\tablefootmark{a} & S36/S37 & 1454 & 20:32:21.850 & +40:20:00.708 \\
 & S38 & 2210 & 20:32:22.302 & +40:19:19.524 \\ 
\hline
33\tablefootmark{a} & N36/N40/N41 & 1018 & 20:38:59.333 & +42:23:37.183 \\
 & N37/N43 & 5417 & 20:38:58.299 & +42:24:35.896 \\
 & N38/N48 & 699 & 20:39:00.023 & +42:22:16.036 \\
 & N44 & 1467 & 20:38:59.642 & +42:23:06.864 \\
 & N51 & 1243 & 20:39:02.409 & +42:25:09.124 \\
 & N52/N53 & 1599 & 20:39:03.131 & +42:26:00.013 \\ 
\hline
\end{tabular}
}
\tablefoot{\\
\tablefoottext{a}{Mosaic field}\\
\tablefoottext{b}{MDCs from \citet{2007A&A...476.1243M}}\\
\tablefoottext{c}{MDCs from \citet{2021ApJ...918L...4C} covered in our SMA field}\\
}
\end{table*}

\begin{figure}[!ht]
    \centering
    \includegraphics[width=17cm]{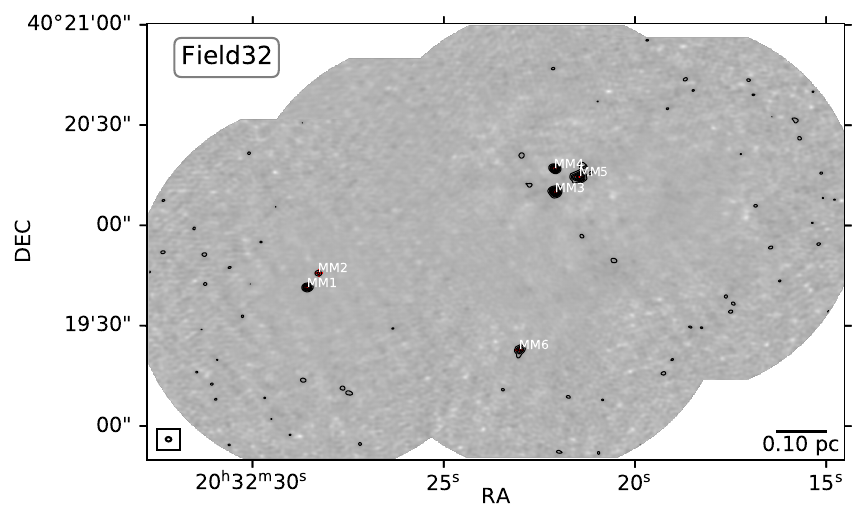}
    \caption{SMA 1.3 mm continuum map of DR15. Similar to Fig \ref{fig:1.3mmContMap001}}
    \label{fig:1.3mmContMap003}
\end{figure}

\begin{figure}[!ht]
    \centering
    \includegraphics[width=17cm]{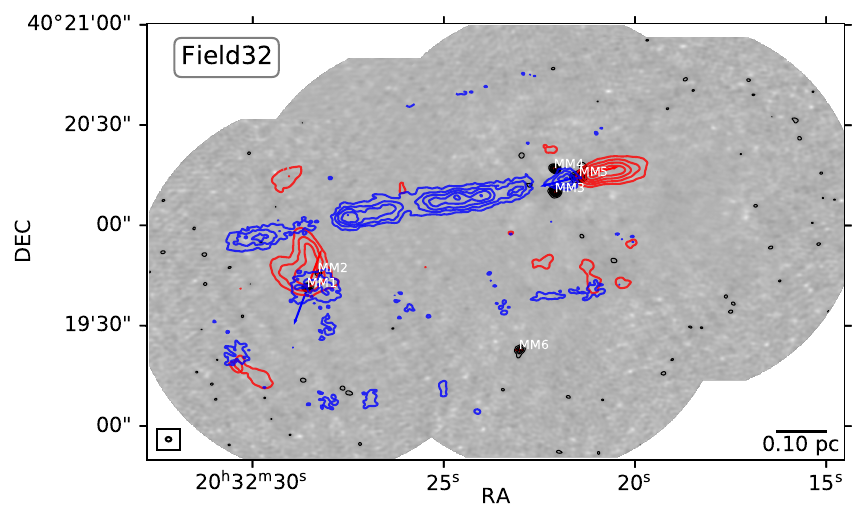}
    \caption{SMA CO outflow maps of DR15. Similar to Fig \ref{fig:OutflowOverCont001}}
    \label{fig:OutflowOverContdr15}
\end{figure}

\begin{figure}[!ht]
    \centering
    \includegraphics[width=11cm]{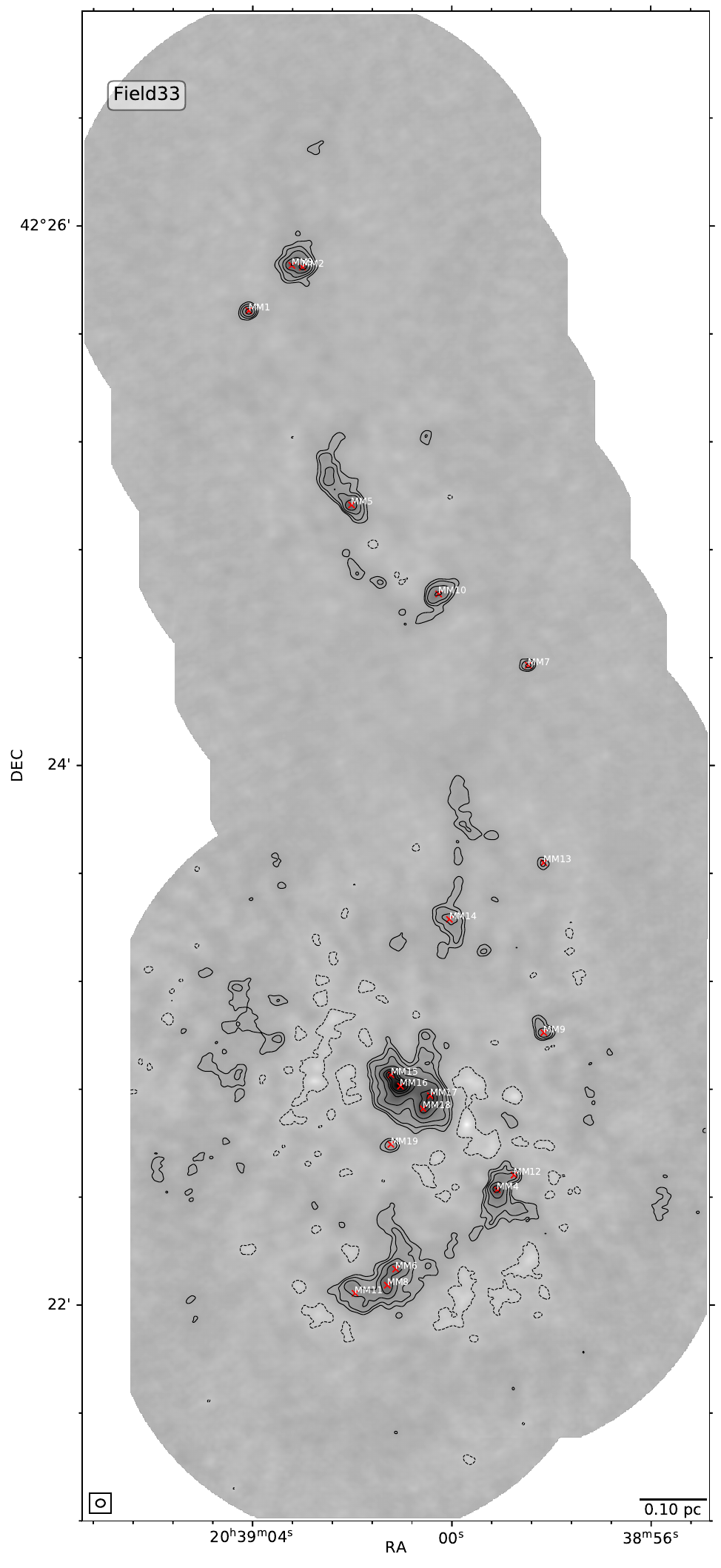}
    \caption{SMA 1.3 mm continuum map of DR21OH. Similar to Fig \ref{fig:1.3mmContMap001}}
    \label{fig:1.3mmContMap004}
\end{figure}
\newpage
\begin{figure}[!ht]
    \centering
    \includegraphics[width=11cm]{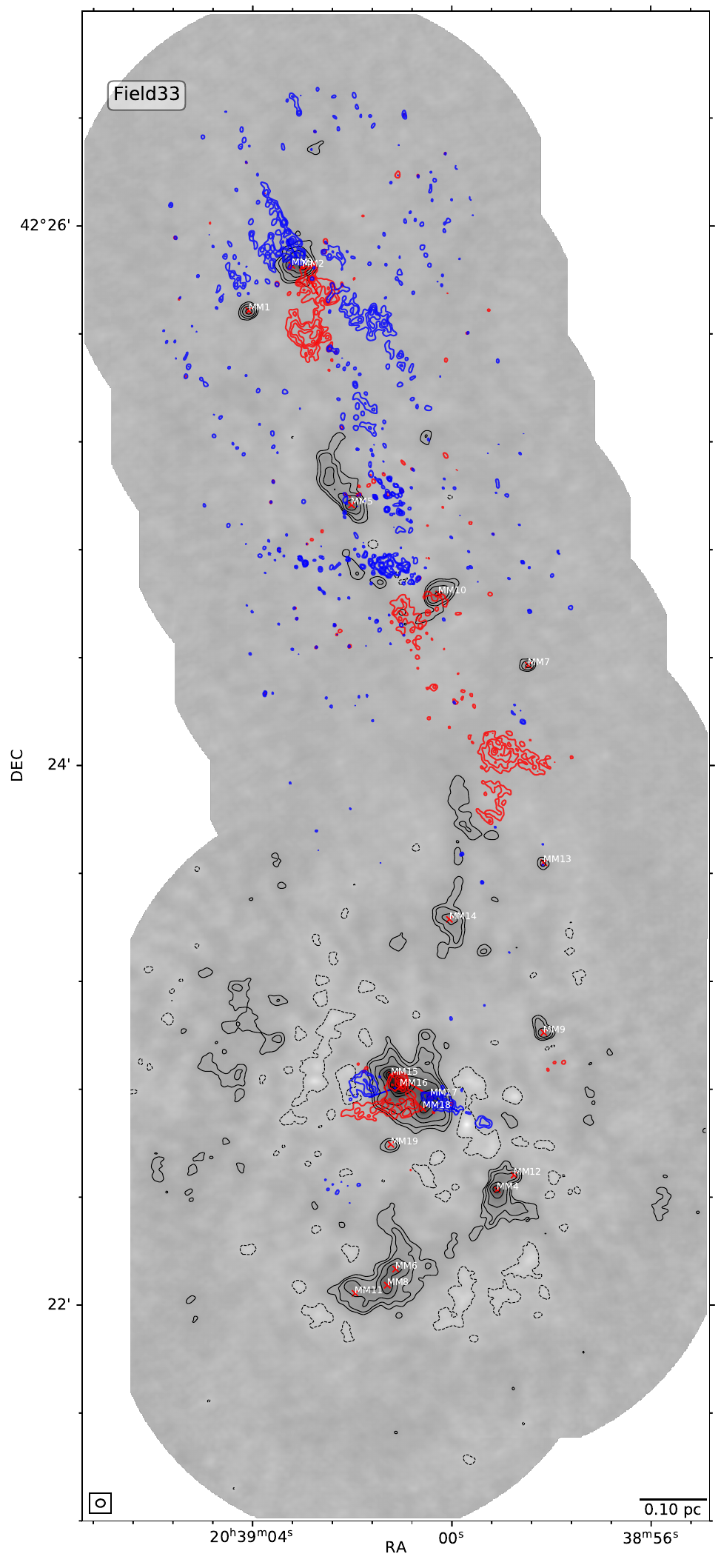}
    \caption{SMA CO outflow maps of DR21OH. Similar to Fig \ref{fig:OutflowOverCont001}}
    \label{fig:OutflowOverContDR21OH}
\end{figure}
\clearpage

\section{Gas temperature} \label{appe:temp}
\begin{figure*}[!h]
    \centering
    \includegraphics[width=10cm]{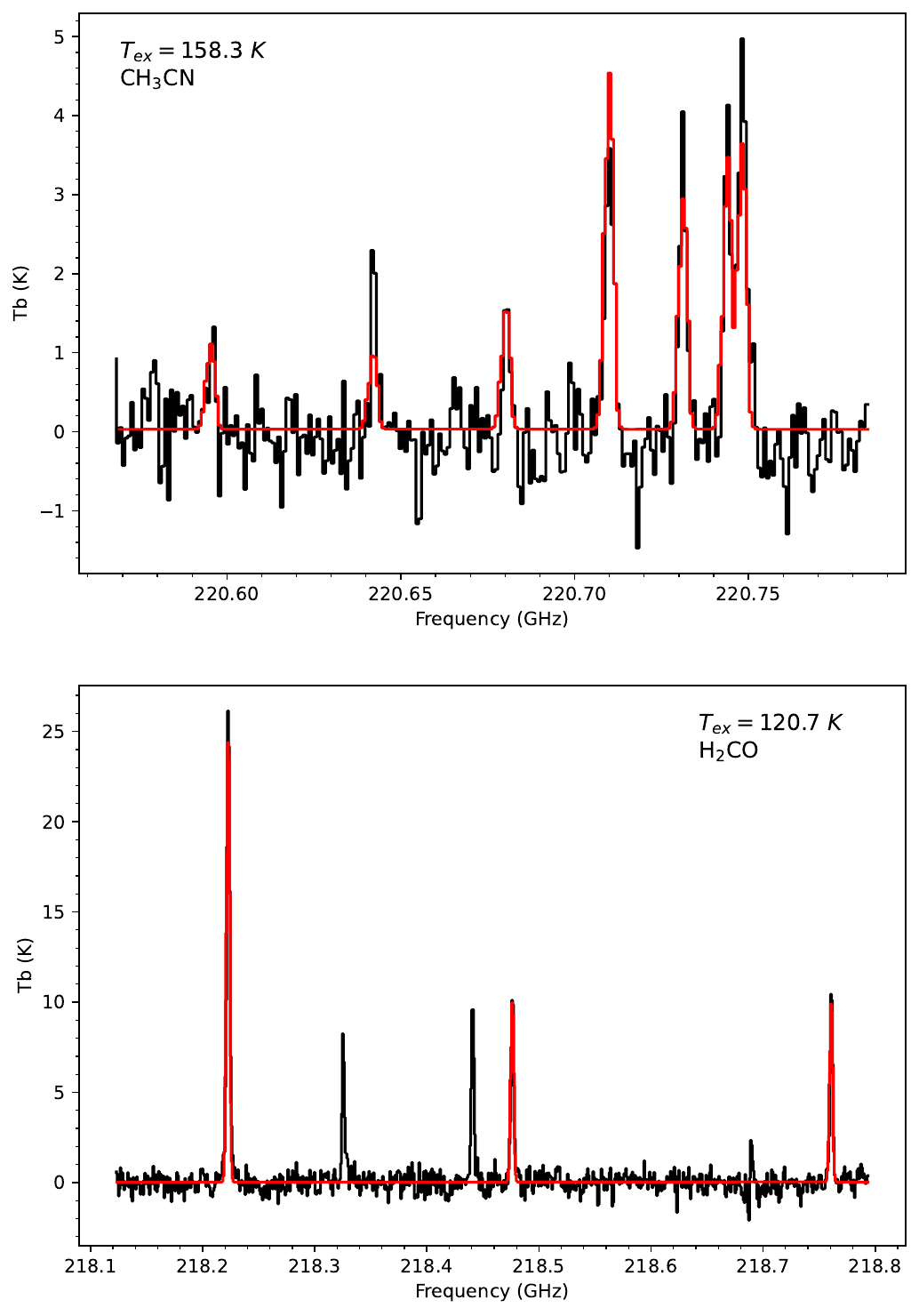}
    \caption{XCLASS fitting results of $\mathrm{CH_3CN}$ and $\mathrm{H_2CO}$ toward Field5 MM1. The black lines represent line emission averaged in the FWHM of the condensation. The red lines are the fitting results from XCLASS. The derived gas temperature from XCLASS is labeled in the map.}
    \label{fig:xclass_field5mm1}
\end{figure*}

We detected several components of $\mathrm{CH_3CN(12_k-11_k)}$ k-ladder (listed in Table \ref{tab:lineinfo}) in a few sources in our sample. The $\mathrm{CH_3CN(12_k-11_k)}$ level is helpful in temperature estimates. The three transitions of $\mathrm{H_2CO}$ with rest frequencies of 218.222, 218.475, and 218.760 GHz can also used for estimations of temperature since they are strong and can be fitted simultaneously within a single spectral window in our SMA observations. We have carefully checked the potential influence of absorption and outflow on temperature estimates with $\mathrm{H_2CO}$ emission. We used the XCLASS (eXtend CASA Line Analysis Software Suite) tool \citep{2017A&A...598A...7M} to fit the lines for temperature estimates. The myXCLASS function in XCLASS models a spectrum by solving the radiative transfer equation for an isothermal object in one dimension. For example, Fig \ref{fig:xclass_field5mm1} shows the fitting results of $\mathrm{CH_3CN(12_k-11_k)}$ and $\mathrm{H_2CO}$ in Field5 MM1. The derived gas temperatures are listed in Table \ref{tab:tgas}. Sources that have $\mathrm{H_2CO}$ emission affected by outflows are marked by asterisks. It is worth noting that the gas temperatures estimated from transitions of $\mathrm{CH_3CN}$ and $\mathrm{H_2CO}$ are higher than the dust temperature from \citet{2021ApJ...918L...4C}. Because of our coarse resolution, the emission of dense gas tracers, especially $\mathrm{CH_3CN}$, probably mainly comes from the inner, unresolved structure surrounding the protostar, which is denser and hotter. It could make the gas temperatures derived from these dense gas tracers much higher than the dust temperatures.

Table \ref{tab:tgas} also includes $\mathrm{NH_3}$ gas temperatures from \citet{zhang2023subm}. Their VLA 1.3 cm observations on Cygnus-X cover some MDCs in our SMA observations. They used the lowest NH3 transitions, $\mathrm{NH_3}$ (1, 1) and (2, 2), to estimate the rotational temperature. Their angular resolution was sufficient (about 3'') to distinguish two individual condensations.

It is important to note that only a few condensations associated with outflow possess multiple transitions of $\mathrm{CH_3CN}$ and $\mathrm{H_2CO}$ for the temperature estimate and many sources have no $\mathrm{NH_3}$ gas temperature from \citet{zhang2023subm}. Consequently, a comprehensive comparison of gas temperatures between disks and non-disks, like what we did with dust temperature, cannot be achieved.

\begin{table*}[!ht]
\centering
\caption{Estimation of molecular gas temperature}
\label{tab:tgas}
{
\begin{tabular}{lllllll}
\hline
\hline
Condensation & $T_\mathrm{CH_3CN}$\tablefootmark{a} & ${T_\mathrm{H_2CO}}$\tablefootmark{a} & $T_\mathrm{NH_3}$\tablefootmark{b} & $T_\mathrm{d}$ & Group \\
& (K) & (K) & (K) & (K) &   \\
\hline
Field3 MM1 & ... & 94\tablefootmark{*} & ... & 40 & Disk Candidate \\
Field4 MM1 & 136 & 194\tablefootmark{*} & 38  & 37 & Disk Candidate \\
Field5 MM1 & 158 & 121\tablefootmark{*} & 31  & 31 & Disk Candidate     \\
Field7 MM2 & 182 & 111\tablefootmark{*} & 110 & 35 & Disk Candidate     \\
Field8 MM1 & ... & 88\tablefootmark{*}  & 25  & 40 & Disk Candidate     \\
Field9 MM1 & 172 & 147\tablefootmark{*}   & ... & 17 & Disk Candidate     \\
Field16 MM1 & 197 & 364\tablefootmark{*} & ... & 32 & Disk Candidate     \\
Field18 MM1 & ... & ... & 27  & 27 & Disk Candidate   \\
Field21 MM1 & ... & 121\tablefootmark{*} & 24  & 34 & Disk Candidate \\
Field1 MM1  & ... & 63\tablefootmark{*}  & ... & 18 & Dense-Gas-Traced \\
Field1 MM2 & ... & 74\tablefootmark{*}  & 40  & 25 & Dense-Gas-Traced     \\
Field4 MM3 & ... & 173\tablefootmark{*}  & 30  & 18 & Dense-Gas-Traced   \\
Field10 MM3 & ... & ...  & 20  & 18 & Dense-Gas-Traced \\
Field11 MM1 & ... & 103\tablefootmark{*} & 33  & 37 & Dense-Gas-Traced \\
Field12 MM1 & ... & ...  & ...  & 22 & Dense-Gas-Traced \\
Field14 MM1 & ... & ... & ...  & 42 & Dense-Gas-Traced \\
Field17 MM3 & ... & 69\tablefootmark{*} & 20  & 21 & Dense-Gas-Traced   \\
Field22 MM2 & ... & 70\tablefootmark{*} & 17  & 18 & Dense-Gas-Traced   \\
Field23 MM1 & ... & ... & 14  & 26 & Dense-Gas-Traced \\
Field26 MM1 & ... & ... & ...  & 21 & Dense-Gas-Traced \\
Field29 MM2 & ... & ... & ...  & 23 & Dense-Gas-Traced \\
Field32 MM1 & ... & ... & ...  & 43 & Dense-Gas-Traced \\
Field32 MM5 & ... & ... & ...  & 16 & Dense-Gas-Traced \\
Field33 MM2 & ... & 15\tablefootmark{*} & ... & 37 & Dense-Gas-Traced \\
Field33 MM3 & ... & 16\tablefootmark{*} & ... & 18 & Dense-Gas-Traced \\
Field33 MM16 & ... & 280\tablefootmark{*}  & ... & 35 & Dense-Gas-Traced \\
Field33 MM17 & ... & 18\tablefootmark{*} & ...  & 27 & Dense-Gas-Traced \\
Field2 MM4  & ... & ...   & ... & 42 & Outflow-Associated \\
Field3 MM2 & ... & ... & 21  & 24 & Outflow-Associated \\
Field6 MM2 & ... & ...  & 17  & 16 & Outflow-Associated \\
Field8 MM3 & ... & ... & ... & 17 & Outflow-Associated \\
Field10 MM2 & ... & ... & ... & 23 & Outflow-Associated \\
Field10 MM4 & ... & ... & ... & 26 & Outflow-Associated \\
Fiedl11 MM3 & ... & ... & ... & 23 & Outflow-Associated \\
Field14 MM2 & ... & ... & 16  & 18 & Outflow-Associated \\
Field17 MM4 & ... & ...  & ... & 19 & Outflow-Associated \\
Field17 MM5 & ... & ...  & ... & 22 & Outflow-Associated \\
Field17 MM8 & ... & ...  & 18  & 42 & Outflow-Associated \\
Field18 MM3 & ... & ...  & 15  & 20 & Outflow-Associated \\
Field20 MM2 & ... & ...  & ... & 22 & Outflow-Associated \\
Field21 MM3 & ... & ...  & 24  & 30 & Outflow-Associated \\
Field21 MM4 & ... & ...  & ... & 35 & Outflow-Associated \\
Field21 MM7 & ... & ...  & ... & 38 & Outflow-Associated \\
Field22 MM3 & ... & ...  & ... & 18 & Outflow-Associated \\
Field25 MM3 & ... & ...  & ... & 15 & Outflow-Associated \\
Field26 MM2 & ... & ...  & ... & 21 & Outflow-Associated \\
Field28 MM1 & ... & ...  & ... & 30 & Outflow-Associated \\
Field28 MM2 & ... & ...  & ... & 28 & Outflow-Associated \\
Field31 MM1 & ... & ...  & ... & 21 & Outflow-Associated \\
\hline
\end{tabular}
}
\tablefoot{\\
\tablefoottext{a}{Gas temperature derived by XCLASS.}\\
\tablefoottext{b}{Ammonia temperature from \citet{zhang2023subm}.}\\
\tablefoottext{*}{$\mathrm{H_2CO}$ emission affected by outflows.}
}
\end{table*}




\end{appendix}

\end{document}